\documentclass[fleqn,usenatbib]{mnras}
\usepackage{footmisc}
\usepackage{graphicx}
\usepackage{amssymb}
\usepackage{amsmath}
\usepackage{color}
\usepackage{psfrag}
\usepackage[caption=false]{subfig}
\usepackage{array}

\newcommand{\oi}{\hbox{ASASSN-15oi}}
\newcommand{\host}{\hbox{2MASX J20390918-3045201}}

\newcommand{\msun}{\hbox{M$_{\odot}$}}

\newcommand{\swift}{\textit{Swift}}

\newcounter{minirefcount}
\def \s{\hphantom{1}}

\usepackage{tikz}

\title[The Late-Time Evolution of ASASSN-15oi]{The Unusual Late-Time Evolution of the Tidal Disruption Event ASASSN-15oi}

\author[T. W.-S. Holoien et al.]{T. W.-S. Holoien,$^{1}$\thanks{E-mail: tholoien@carnegiescience.edu}
J. S. Brown,$^{2}$
K. Auchettl,$^{3,4}$
C. S. Kochanek,$^{2,3}$ \newauthor
J. L. Prieto,$^{5,6}$
B. J. Shappee,$^{7}$
J. Van Saders$^{7}$
\\
\\
$^{1}$ The Observatories of the Carnegie Institution for Science, 813 Santa Barbara St., Pasadena,
CA 91101, USA\\
$^{2}$ Department of Astronomy, The Ohio State University, 140 West 18th Avenue, Columbus, OH 43210, USA\\
$^{3}$ Center for Cosmology and Astro-Particle Physics, The Ohio State University, 191 West Woodruff Avenue, Columbus, OH 43210, USA\\
$^{4}$ Department of Physics, The Ohio State University, 191 W. Woodruff Avenue, Columbus, OH 43210, USA\\
$^{5}$ N\'ucleo de Astronom\'ia de la Facultad de Ingenier\'ia, Universidad Diego Portales, Av. Ej\'ercito 441, Santiago, Chile \\
$^{6}$ Millennium Institute of Astrophysics, Santiago, Chile \\
$^{7}$ Institute for Astronomy, University of Hawai'i, 2680 Woodlawn Drive, Honolulu, HI 96822, USA
}
\date{Accepted in MNRAS, 480, 5869.}

\pubyear{2018}
\begin{document}
\label{firstpage}
\pagerange{\pageref{firstpage}--\pageref{lastpage}}
\maketitle

\begin{abstract}
We present late-time optical spectroscopy and X-ray, UV, and optical photometry of the nearby ($d=214$ Mpc, $z=0.0479$) tidal disruption event (TDE) \oi. The optical spectra span 450 days after discovery and show little remaining transient emission or evolution after roughly 3 months. In contrast, the {\swift} and \textit{XMM-Newton} observations indicate the presence of evolving X-ray emission and lingering thermal UV emission that is still present 600 days after discovery. The thermal component of the X-ray emission shows a unique, slow brightening by roughly an order of magnitude to become the dominant source of emission from the TDE at later times, while the hard component of the X-ray emission remains weak and relatively constant throughout the flare. The TDE radiated $(1.32\pm0.06)\times10^{51}$~ergs across all wavelengths, and the UV and optical emission is consistent with a power law decline and potentially indicative of a late-time shift in the power-law index that could be caused by a transition in the dominant emission mechanism. 
\end{abstract}

\begin{keywords}
accretion, accretion disks -- black hole physics -- galaxies: nuclei
\end{keywords}

\section{Introduction}
\label{sec:intro}
When a star passes within the tidal radius of a supermassive black hole (SMBH), the tidal shear forces overwhelm the self-gravity of the star, tearing it apart. If the star is on an approximately parabolic orbit, roughly half of the stellar material will be ejected, while the other half will remain bound and return to pericenter at a rate proportional to $t^{-5/3}$, resulting in a luminous accretion flare \citep{rees88,evans89,phinney89}. These tidal disruption events (TDEs) exhibit a wide range of observational properties that depend on many factors, including the physical properties of the disrupted star \citep[e.g.,][]{macleod12,kochanek16}, the post-disruption evolution of the accretion stream \citep[e.g.,][] {kochanek94,strubbe09,guillochon13,hayasaki13,hayasaki16,piran15,shiokawa15}, and complex radiative transfer effects \citep[e.g.,][]{gaskell14,strubbe15,roth16,roth18}.

If we want to fully understand the observational characteristics and physical properties of these highly energetic events, a large sample of well-studied TDEs is required. While the number of well-studied TDE candidates is growing \citep[e.g.,][]{velzen11,cenko12a,gezari12b,arcavi14, chornock14,holoien14b,gezari15,vinko15,holoien16a,holoien16b,brown16a,blagorodnova17,brown17a,brown17b,hung17}, these candidates show a surprising diversity in observed properties, and the number of TDEs that have been well studied at very early and very late times remains quite small. 

ASASSN-15oi is a nearby ($d=214$ Mpc, $z=0.0479$) TDE discovered by the All-Sky Automated Survey for SuperNovae \citep[ASAS-SN;][]{shappee14} on 2015 August 14 \citep{asassn15oi_atel}. We began an immediate follow-up campaign, observing \oi\ for roughly 3.5 months after discovery using a wide range of both ground and space-based observatories \citep{holoien16b}. These observations showed that \oi\ had spectral features resembling the ``Helium-rich'' TDEs from \citet{arcavi14} and optical/UV evolution that was consistent with that of a blackbody and a luminosity declining at a rate best-fit by an exponential profile (though a power-law profile was also a reasonable fit to the data). Interestingly, while the spectral features had faded almost completely only $\sim3$ months after discovery, the UV emission remained brighter than the host for the entire period of observations presented in \citet{holoien16b}.


\begin{table}
\centering
\caption{Archival Host Galaxy Magnitudes}
\label{table:host_mags1}
\begin{tabular}{@{}ccc}
\hline
Filter & Magnitude & Magnitude Uncertainty \\
\hline
$NUV$ & $>$22.98\s & --- \\
$g$ & 17.66 & 0.01 \\
$r$ & 16.95 & 0.02 \\
$i$ & 16.64 & 0.01 \\
$z$ & 16.45 & 0.01 \\
$J$ & 16.07 & 0.07 \\
$H$ & 15.89 & 0.07 \\
$K_S$ & 15.89 & 0.09 \\
$W1$ & 16.88 & 0.03 \\
$W2$ & 17.46 & 0.05 \\
\hline
\end{tabular}
\\
\medskip
\raggedright
\noindent 
Archival magnitudes of the host galaxy from GALEX ($NUV$), Pan-STARRS ($griz$), 2MASS ($JHK_S$) and WISE ($W1~W2$) that were used for fitting the host galaxy SED. The GALEX $NUV$ limit is a $3$-$\sigma$ upper limit measured from co-added exposures totaling $\sim200$s of exposure time. The Pan-STARRS magnitudes are Kron magnitudes from the Pan-STARRS Data Release 1 catalog \citep{flewelling16}, the 2MASS magnitudes are 5\farcs{0}-radius aperture magnitudes, and the WISE $W1$ and $W2$ are PSF photometry magnitudes from the AllWISE source catalog. All magnitudes are presented in the AB system.
\end{table}

As was the case with the TDE ASASSN-14li \citep{holoien16a,brown17a}, we detected soft X-ray emission from \oi\ during our initial follow-up campaign. However, \oi\ exhibits much weaker emission than ASASSN-14li, and the detected flux was below an archival upper limit from the ROSAT All-Sky Survey \citep{voges99}, making it difficult to ascertain the source of the X-rays, particularly since the flux displayed little evolution during our initial follow-up observations.

Recently \citet{gezari17} presented UV, optical, and X-ray observations obtained with the \swift\ UltraViolet and Optical Telescope \citep[UVOT;][]{roming05} and X-ray Telescope \citep[XRT;][]{burrows05} and the \textit{XMM-Newton} Observatory, finding that the X-ray emission from \oi\ exhibited a delayed brightening roughly a year after discovery. This behavior is unique among optically discovered TDEs that emit X-rays, and \citet{gezari17} argue that the timescale of the brightening suggests that the X-ray emission may be coming from inefficient circularization of the stellar debris stream, resulting in delayed accretion.

In this manuscript we present extended observations of \oi\ going out to $\sim600$ days after discovery, including late-time optical spectra taken with the Inamori-Magellan Areal Camera and Spectrograph \citep[IMACS;][]{dressler11} on the 6.5-meter Magellan-Baade telescope. We also present additional analysis of the extensive UV, optical, and X-ray observations from \swift\ and \textit{XMM-Newton}. In Section~\ref{sec:obs} we describe our observations obtained during this extended monitoring campaign. In Section~\ref{sec:analysis} we analyze these data and model the late-time evolution of \oi. Finally, in Section ~\ref{sec:disc} we summarize our results and discuss the implications for future studies of TDEs.


\section{Observations}
\label{sec:obs}

This section summarizes the optical, UV, and X-ray observations taken during our $\sim600$ day follow-up campaign of \oi.


\subsection{Re-examination of the Host Galaxy \host}

\label{sec:host}

Since the publication of \citet{holoien16b}, Pan-STARRS \citep{chambers16,flewelling16} imaging data of the host galaxy \host\ have become available and provide critical pre-explosion optical fluxes for the galaxy. In this Section we present an updated analysis of the host galaxy \host\ that incorporates archival observations from the Galaxy Evolution Explorer (GALEX), Pan-STARRS, the Two-Micron All Sky Survey \citep[2MASS;][]{skrutskie06}, and the Wide-field Infrared Survey Explorer \citep[WISE;][]{wright10}. The archival magnitude measurements used for this fit are shown in Table~\ref{table:host_mags1}.

We used the publicly available Fitting and Assessment of Synthetic Templates \citep[\textsc{fast};~][]{kriek09} to fit the spectral energy distribution (SED) of the host galaxy. We assumed a \citet{cardelli89} extinction law with $R_V=3.1$ and a Galactic extinction of $A_V = 0.19$ mag \citep{schlafly11}. We employed an exponentially declining star-formation history, a Salpeter initial mass function, and the \citet{bruzual03} stellar population models. The physical parameters of the host galaxy are largely consistent with our previous results: $M_{\star}=1.0^{+0.2}_{-0.1} \times 10^{10}$ M$_{\odot}$, age $=2.2^{+0.8}_{-0.3}$ Gyr, and a star formation rate SFR $\leq 0.002$ M$_{\odot}$~yr$^{-1}$. Our late-time, high signal-to-noise spectra obtained with IMACS are consistent with these properties. In particular, there is no evidence for any nebular emission features that would be indicative of star formation activity.


\begin{table}
\centering
\caption{Synthetic Host Galaxy Magnitudes}
\label{table:host_mags2}
\begin{tabular}{@{}ccc}
\hline
Filter & Magnitude & Magnitude Uncertainty \\
\hline
$UVW2$ & 23.32 & 0.11 \\
$UVM2$ & 22.99 & 0.17 \\
$UVW1$ & 21.63 & 0.09 \\
$U_{UVOT}$ & 19.52 & 0.04 \\
$B_{UVOT}$ & 18.18 & 0.03 \\
$B_{LCO}$ & 18.13 & 0.03 \\
$V_{UVOT}$ & 17.31 & 0.02 \\
$V_{LCO}$ & 17.26 & 0.02 \\
$I$ & 16.54 & 0.01 \\
\hline
\end{tabular}
\\
\medskip
\raggedright
\noindent 
Median synthesized host magnitudes of the host galaxy in the {\swift} UVOT and Las Cumbres Observatory $BVI$ filters and their 68\% confidence intervals. Magnitudes were synthesized using the process described in \S\ref{sec:host} and are presented in the AB system.
\end{table}


\begin{table*}
\centering
\caption{ASASSN-15oi X-ray properties.}
\label{table:xrayprop}
\begin{tabular}{lccccccc}
\hline
 & & Exposure & & & Blackbody & Unabs. flux & Unabs. flux \\
Observation & MJD & Time & kT & $\Gamma$ & Radius & of blackbody  & of powerlaw \\
& s & ks & keV&  & $10^{12}$ cm & $10^{-13}$ erg cm$^{-2}$ s$^{-1}$ & $10^{-14}$ erg cm$^{-2}$ s$^{-1}$ \\
\hline
Swift 001-012 &57275.1 & 24.5 & 0.036$\pm$0.01&$1.14_{-1.1}^{+1.9}$& $1.33\pm0.73$ & $ 1.1^{+0.6}_{-0.6}$ & $1.6^{+1.3}_{-1.3} $\\
XMM 0722160501 & 57324.6 & 12.4 & 0.062$\pm$0.06 & $1.68_{-0.8}^{+1.0}$& $0.47\pm0.60$ & $ 1.2^{+0.5}_{-0.5}$ & $2.3^{+0.8}_{-0.8} $\\
Swift 013-027& 57325.2 & 36.4 &0.058$\pm$0.01 &$1.49_{-1.4}^{+1.9}$& $0.47\pm0.16$ & $ 0.8^{+0.3}_{-0.3}$ & $1.6^{+1.2}_{-1.2} $\\
Swift 028-031 &57474.6& 12.9 & 0.047$\pm$0.01&$\dotsm$ & $2.03\pm0.70$ & $ 6.8^{+1.6}_{-1.6}$ & $<9.1$\\
XMM 0722160701 & 57482.7 & 14.0 & 0.053$\pm$0.02 & $2.78_{-1.2}^{+2.6}$ & $1.08\pm0.82$ & $ 3.2^{+0.9}_{-1.0}$ & $1.5^{+0.8}_{-0.8} $\\
Swift 032-045 & 57545.1& 27.4 & 0.049$\pm$0.01&$\dotsm$  & $2.01\pm0.57$ & $ 8.2^{+1.4}_{-1.1}$ & $<8.1$\\
Swift 046-050 & 57607.4& \s3.9 & 0.047$\pm$0.02&$1.88^{+3.6}_{-1.0} $& $2.27\pm1.94$ & $ 8.6^{+3.3}_{-3.3}$ & $9.0^{+7.0}_{-1.4} $\\
Swift 052-053 &57839.3 & \s4.1 &0.086$\pm$0.08& $\dotsm$ & $0.23\pm0.46$ & $ 1.0^{+0.9}_{-0.9}$ & $<87$\\
\hline
\end{tabular}
\\
\medskip
\raggedright
\noindent 
X-ray properties of \oi\ measured from the \swift\ XRT and \textit{XMM-Newton} observations. \swift\ data were binned in time to increase the signal-to-noise of the observations; epochs combined in each bin are indicated in Column 1 and total exposure time is given in Column 3.
\end{table*}

With the improved constraints on the SED of the host galaxy, we re-derived the synthetic host magnitudes for each photometric band in our follow-up campaign. In order to determine a robust estimate of the host SED (and uncertainties) we generated 1000 realizations of the archival fluxes, perturbed by their respective uncertainties and assuming Gaussian errors. We then modeled each realization and computed the synthetic magnitudes for each resulting SED. For each filter this yields a distribution of synthetic magnitudes; we report the median and 68\% confidence intervals on the host magnitudes in Table~\ref{table:host_mags2}.

The host magnitudes differ slightly from those derived in our initial analysis; most importantly, we find the host to be roughly 0.1 magnitudes fainter in the {\swift} UV filters than previously estimated. Given the luminosity of the flare, this has little impact on the early-time properties of the flare, but the improved constraints on the host galaxy are crucial for properly interpreting the nature of the flare at later times when the flare has faded considerably.


\subsection{Spectroscopic Observations}
\label{sec:spec_obs}

In addition to the spectra presented in \citet{holoien16b}, which spanned from 2015 August 21 and 2015 November 7, we obtained low-resolution optical spectra of ASASSN-15oi with IMACS on the Magellan Baade 6.5m telescope and with the Wide-Field Reimaging CCD camera and spectrograph (WFCCD) on the du Pont 2.5m telescope at Las Campanas Observatory in Chile.

The spectra were obtained on UT 2016 Jun. 10.2 (WFCCD, $3\times 1200$~sec), 2016 Aug. 3.1 (IMACS, $5\times 1450$~sec), and 2016 Nov. 6.0 (IMACS, $2\times 1200$~sec) using a 0\farcs{7} slit and the 300~l/mm grism with f/2 camera on IMACS and 1\farcs{7} slit with 400~l/mm blue grism on WFCCD which yielded spectra with a spectral resolution of $R= 700-1300$. We reduced and extracted the spectra using standard routines in \textsc{iraf}, including bias subtraction, flat fielding and 1D spectral extraction. The spectra were calibrated in wavelength using HeNeAr lamps obtained at the time of the science exposures and the flux calibration was obtained using the observation of a spectrophotometric standard observed the same night. We discuss the spectral evolution of \oi\ in Section~\ref{sec:spec_anal}.


\subsection{\swift\ Observations}
\label{sec:swift_obs}

We also obtained and analyzed additional publicly available data from \swift\ from 2016 March 21 through 2017 April 2 (\swift\ target ID 33999; PIs: Hallefors, Horesh, Holoien, Cenko, Gezari). The UVOT observations were obtained in the $V$ (5468 \AA), $B$ (4392 \AA), $U$ (3465 \AA), $UVW1$ (2600 \AA), $UVM2$ (2246 \AA), and $UVW2$ (1928 \AA) filters \citep{poole08}. As in \citet{holoien16b}, we used the UVOT software task \textsc{uvotsource} to extract the source counts from a 5\farcs0 radius region and a sky region with a radius of $\sim$~40\arcsec. The UVOT count rates were converted into magnitudes and fluxes based on the most recent UVOT calibration \citep{poole08,breeveld10}. We present the UVOT magnitudes in Table~\ref{tab:phot}.

In addition to the UVOT observations, {\oi} was also observed using the photon counting mode of the {\swift} XRT. All observations were reprocessed from level one XRT data using the \emph{Swift} \textsc{xrtpipeline} version 0.13.2 script, as suggested in the \textit{Swift} XRT data reduction guide\footnote{\url{http://swift.gsfc.nasa.gov/analysis/xrt\_swguide\_v1\_2.pdf}}. Due to a bright X-ray source $\sim$40 arcsec away from the position of \oi, we used a source region centered on the position of {\oi} with a radius of 20\farcs0, and a source free-background region centered at  $(\alpha,\delta)=($20:39:25.2, $-$30:45:26.2$)$ with a radius of 110\farcs0 to extract the count rate of this event. All extracted count rates were corrected for the fact that our source region encloses only a fraction of the total number of counts arising from the source. Here, the 20\farcs{0} radius encloses 80\% of the flux from the source \citep{moretti04}. We also find that none of our observation suffer from pileup issues. 


\begin{figure*}
\begin{minipage}{\textwidth}
\centering
\subfloat{{\includegraphics[width=0.95\textwidth]{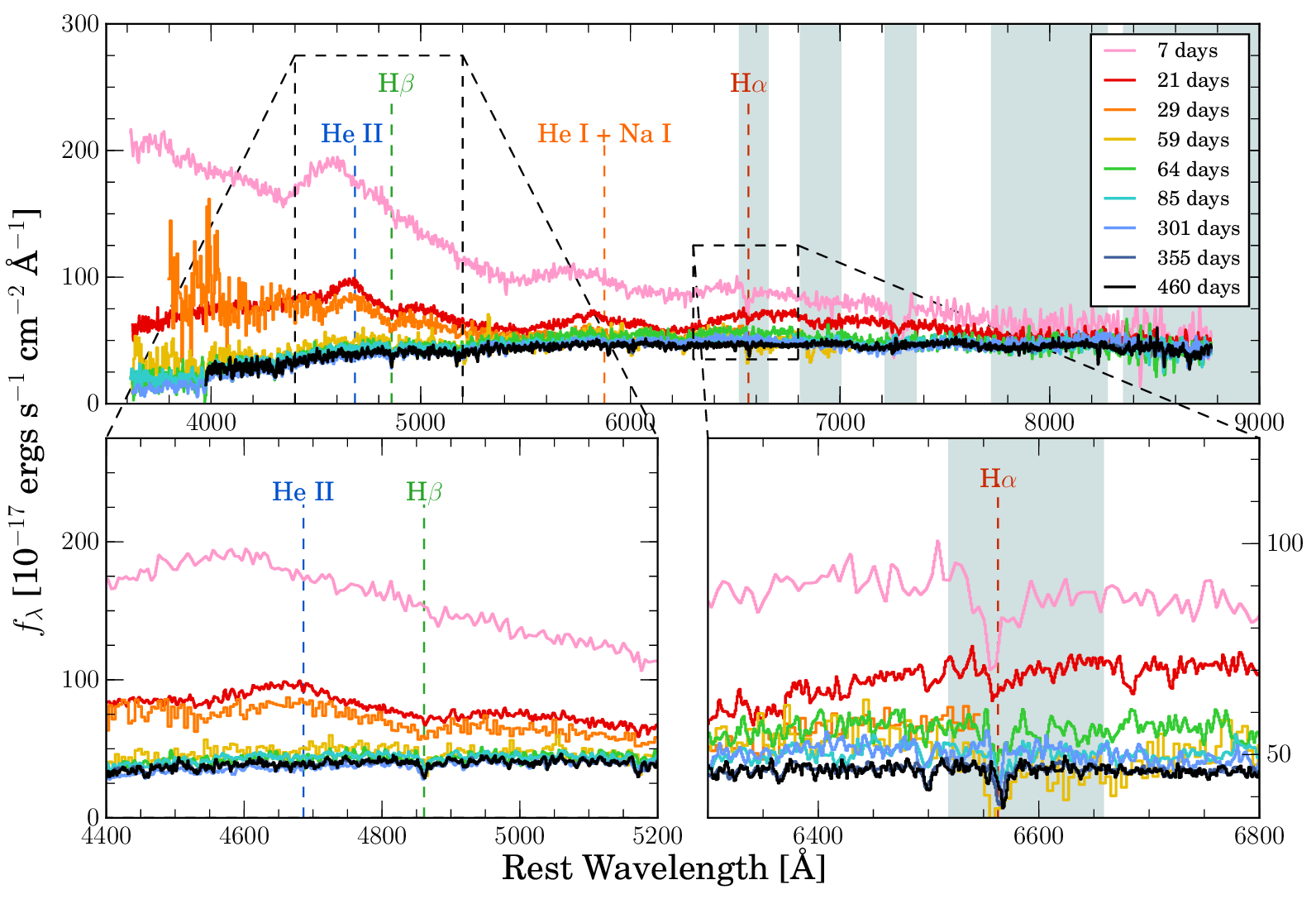}}}
\caption{The evolution of the optical spectra of \oi\ beginning 7 days after discovery (pink) and ending 460 days after discovery (black). The top panel shows the full optical spectra, while the bottom panels show expanded views of the regions around \ion{He}{ii} $\lambda4686$ (left) and H$\alpha$ (right). Prominent spectral features commonly observed in TDEs are labeled, and regions prone to systematic errors related to telluric correction are indicated by the shaded bands.}
\label{fig:spec_evol}
\end{minipage}
\end{figure*}
		
To increase the signal-to-noise of our observations, we also combined the \textit{Swift} observations into six time bins using \textsc{xselect} version 2.4d. From these merged observations, we used the task \textsc{xrtproducts} version 0.4.2 to extract both source and background spectra using the same regions used to extract our count rates. To generate ancillary response files for each spectra, we used the task \textsc{xrtmkarf} and the exposure maps of each observation produced by the \textsc{xrtpipeline} and merged them using \textsc{ximage} version 4.5.1.  The response matrix files are ready-made files which were obtained from the CALDB. 


\subsection{XMM-Newton Observations}
\label{sec:xmm_obs}

ASASSN-15oi was observed with both the MOS and PN detectors on the \textit{XMM-Newton} Observatory on 2015 October 29 and 2016 April 4 for a total of 16\,ks each (ObsID: 0722160501 and 0722160701, PI: Gezari). Both detectors were operated in the full frame mode using a thin filter, however we use only the PN detector for our study due its the high sensitivity, and large effective area. All data reduction and analysis was done using the \textit{XMM-Newton} science system (SAS) version 15.0.0\footnote{\url{https://www.cosmos.esa.int/web/xmm-newton/documentation/}} with CALDB 4.6.8\footnote{\url{https://www.cosmos.esa.int/web/xmm-newton/calibration}}.

We first checked for periods of high background and/or proton flares by generating a count rate histogram using events with energy between 10--12\,keV. We find that these observations are only slightly affected by high background or flares, giving effective exposure times in the PN detector of $\sim11$\,ks and $\sim$13\,ks for 0722160501 and 0722160701, respectively.  We reduced the data following the standard screening of events, with single and double events (PATTERN $\le$ 4) selected for the PN detector. We also used the standard set of FLAGS for the PN (\#XMMEA\_EP) detector. We also corrected for vignetting by processing all event files using the task \textsc{evigweight}\footnote{See \url{https://xmm-tools.cosmos.esa.int/external/sas/current/doc/evigweight} for more details}. We extracted the spectrum of ASASSN-15oi using the SAS task \textsc{evselect} and the cleaned, vignetting-corrected event files from the PN detector using the same source and background region that we used for the {\swift} observations. Count rates are also extracted from the PN observations using the same regions and corrected for the encircled energy fraction\footnote{\url{https://heasarc.nasa.gov/docs/xmm/uhb/onaxisxraypsf.html}}. These data are analyzed in Section~\ref{sec:phot_anal} and Section~\ref{sec:xray_anal}.


\section{Evolution of the Late-Time Emission}
\label{sec:analysis}

In this section we discuss the optical, UV, and X-ray evolution of \oi\ and interpret the physical conditions of the system.


\subsection{Evolution of the Optical Spectra}
\label{sec:spec_anal}


\begin{figure}
\centering
\subfloat{{\includegraphics[width=0.45\textwidth]{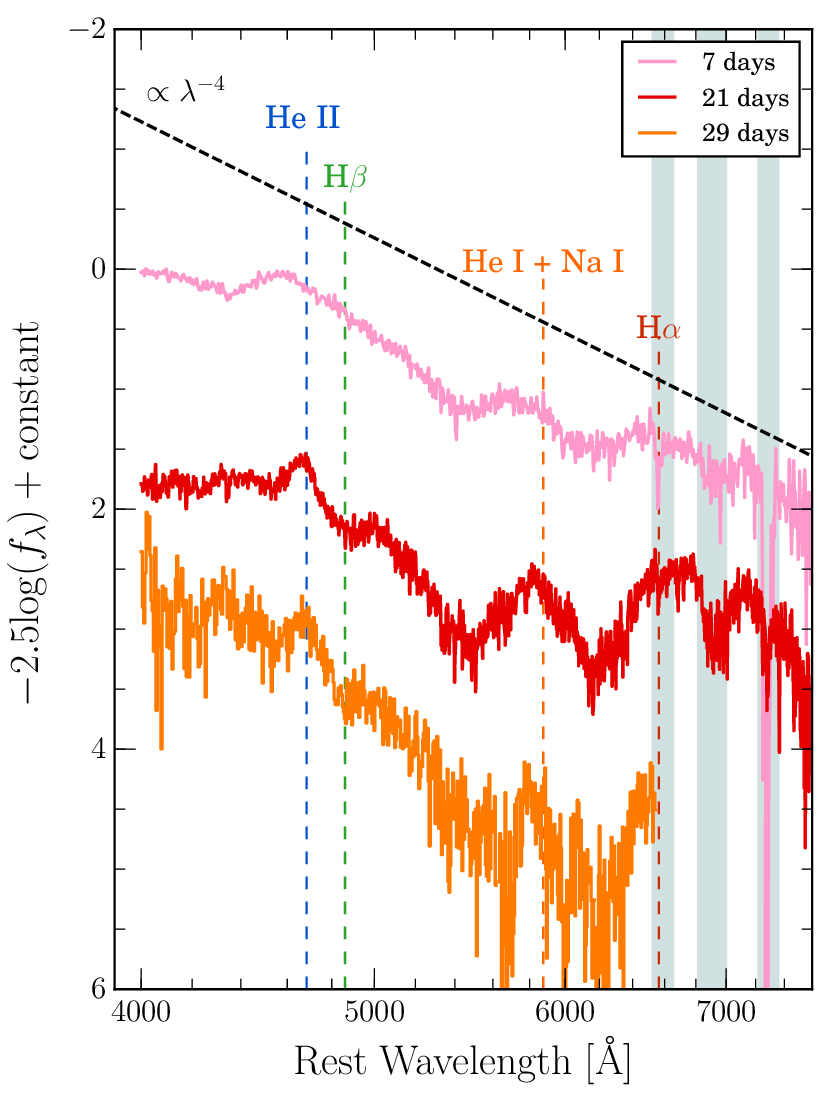}}}
\caption{Host-subtracted early time spectra. The color scheme is the same as Figure~\ref{fig:spec_evol}, and the same prominent TDE emission features are labeled. Note that the absorption feature near H$\alpha$ is due to telluric absorption. The flare continuum is reasonably approximated by the Raleigh-Jeans tail of a blackbody; the host subtracted spectra clearly show the early-time evolution of the emission features.}
\label{fig:spec_sub_early}
\end{figure}

In \citet{holoien16b} we presented spectra spanning the first 85 days after the discovery of \oi, noting that the TDE exhibited a strong blue continuum and broad helium line emission but little-to-no hydrogen emission features in its spectra. The spectra most closely matched those of the ``He-rich'' events from \citet{arcavi14}, such as PS1-10jh and PTF09ge \citep{gezari12b,arcavi14}. As was the case with ASASSN-14ae and ASASSN-14li, the blue continuum weakened and the emission lines became narrower and weakened over time, but unlike the other ASAS-SN TDEs, the transient emission features seemed to fade completely from the spectra within roughly 2 months after discovery.

Figure~\ref{fig:spec_evol} shows the evolution of the optical spectra of \oi\ beginning 7 days after discovery (pink) and ending 460 days after discovery (black). The top panel shows the full optical spectra, while the bottom panels show expanded views of the regions around \ion{He}{ii} $\lambda4686$ (left) and H$\alpha$ (right). After the initial $\sim2$~months, there is surprisingly little evidence of any TDE signatures whatsoever. In the absence of any archival spectra, this was not immediately clear when we initially analyzed this object. However, the high signal-to-noise spectra obtained at late-times with IMACS confirm that the late-time spectra are overwhelmingly dominated by the host galaxy light.

In Figure~\ref{fig:spec_sub_early} we use the late-time, high signal-to-noise spectrum as a template for the host and subtract it from the early-time flare-dominated spectra. The trends seen in the Figure~\ref{fig:spec_evol}, in particular the evolution towards lower velocities and narrower lines, are even more pronounced in these host-subtracted spectra. This evolution takes place on a much shorter timescale than most other TDEs (e.g., ASASSN-14ae; \citealt{holoien14b,brown16a} and ASASSN-14li; \citealt{holoien16a,brown17a}). However, one other nearby TDE, iPTF16fnl, exhibited a similar spectroscopic evolution \citep{blagorodnova17,brown17b}. Both objects displayed rapidly fading optical spectra, though at early times the emission lines in \oi\ are much more blueshifted than the lines seen in iPTF16fnl. The host-subtracted spectra also show that the continuum shape is broadly consistent with the Raleigh-Jeans tail of a blackbody, consistent with the photometric data. 

Finally, the host-subtracted spectra make it quite clear that there is likely some H$\alpha$ emission at early times. While the rest frame H$\alpha$ line emission falls within the B-band telluric feature, there is compelling evidence of a broad blue wing associated with H$\alpha$ emission. Within the context of the broader population of the TDEs, \oi\ remains in the relatively He-rich regime, especially when compared ASASSN-14ae and ASASSN-14li.


\subsection{Analysis of the X-ray Emission}
\label{sec:xray_anal}

To analyze the X-ray spectra of \oi, we used the X-ray spectral fitting package \textsc{xspec} version 12.9.1 using chi-squared statistics. Using the \textsc{ftools} command \textsc{grppha}, we group each spectrum from the merged \textit{Swift} observations and the \textit{XMM-Newton} observations with a minimum of 10 counts per energy bin. Early \textit{Swift} observations of ASSASN-15oi showed that the X-ray emission from this event can be well fit using an absorbed blackbody plus power law \citep{holoien16b}. As we are interested in how the X-ray emission from the source changes with time, we fit each spectra over an energy range of 0.3--5.0\, keV using either an absorbed blackbody model  or an absorbed blackbody plus power law model that is emitting at the redshift the TDE. Here, we assume \citet{wilms00} abundances, and initially fit $N_{H}$, the temperature of the blackbody (kT), the power law index ($\Gamma$), and the normalization. We find that for nearly all spectra, letting $N_{H}$ vary does not significantly improve the fit to the data, and thus we fix it to the Galactic H\textsc{i} column density in the direction of ASASSN-15oi, $5.59\times10^{20}$ cm$^{-2}$  \citep{kalberla05}.

Similar to \citet{holoien16b}, we find that at early times ($<200$ days), the X-ray emission of ASASSN-15oi can be well described using an absorbed blackbody plus power law model. At late times ($>200$ days) we find that having an additional power law component is not required for all \textit{Swift} epochs. Within the uncertainties, there is little evidence for variation in the power law index when a power law component is detected, however at late times the flux of this component seems to increase slightly when compared to the second XMM observation (see Table~\ref{table:xrayprop}). 


\begin{figure}
\centering
\subfloat{{\includegraphics[width=0.45\textwidth]{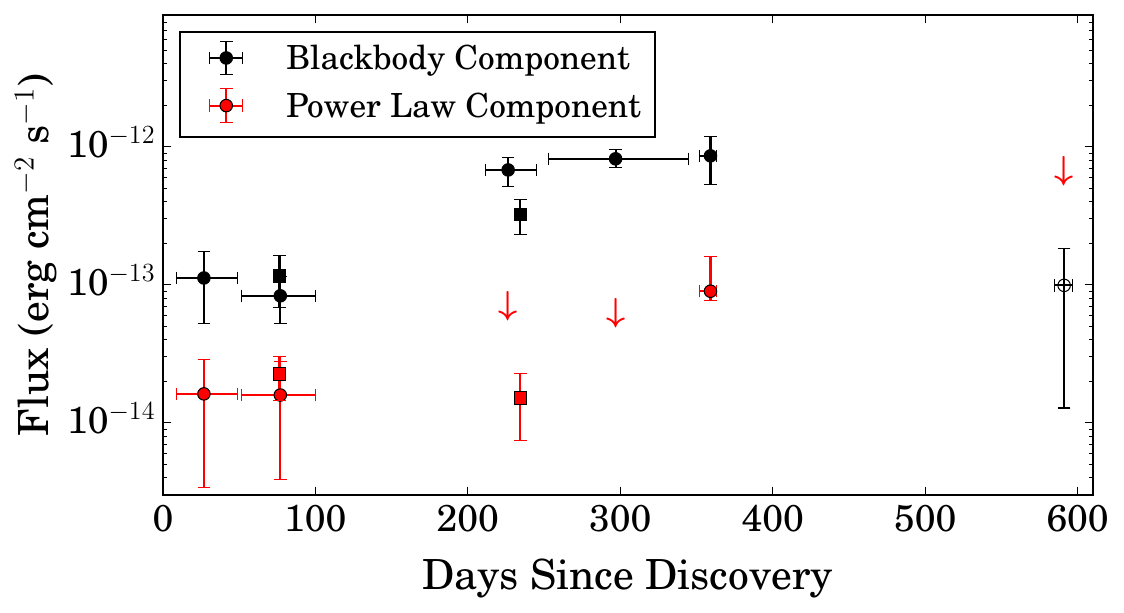}}}
\caption{The flux from the blackbody (black points) and power law (red points) components of the X-ray emission. Arrows indicate upper limits on the flux. In epochs where both components are detected, the blackbody flux is roughly an order of magnitude stronger than the power law flux, and the blackbody component shows a significant long-term brightening between $\sim200$ and $\sim400$ days, while the power law component remains relatively constant.}
\label{fig:xray_flux_comp}
\end{figure}

The bulk of the X-ray emission from ASASSN-15oi can be described using a simple absorbed blackbody model. By merging the available \textit{Swift} observations into six time bins we were able to derive the flux of the blackbody component (and the power law component if present) by modeling the X-ray spectrum. Similar to what was found by \citet{gezari17}, Figure~\ref{fig:xray_flux_comp} shows that the unabsorbed X-ray flux from the blackbody component brightened by a factor of 4--10 between the emission detected at early times ($<200$ days) and that detected at late times ($>200$ days), despite the temperature of the blackbody component remaining constant (within uncertainties) during this time. This behavior is not seen in other X-ray TDEs over such long timescales. Most of the X-ray TDEs, including ASASSN-15oi, exhibit some variability on short timescales. However, over longer timescales the X-ray emission  usually decays following a simple power law (see, e.g., \citet{auchettl17} for the long term behavior of X-ray TDEs, or individual studies such as those of ASASSN-14li: \citealt{brown17a} and Swift J1644+57: \citealt{mangano16} for variability observed over short timescales). This is in contrast with the flux of the power law component, which seems to stay relatively constant with time. \citet{gezari17} suggested that the unique behavior of the thermal component of this source arises from inefficient circularization, resulting in delayed accretion, though this could also be explained by the material surrounding the accretion disk being optically thick to X-ray radiation at early times before becoming optically thin a few months after discovery \citep[e.g.,][]{metzger15}. The former explanation is more likely in this case as we do not see any significant change in $N_H$ from the X-ray observations.

Previously, we placed strong limits on the presence of an underlying AGN based on a non-detection from archival \textit{ROSAT} all-sky survey data \citep{holoien16b}. However, a re-examination of this calculation discovered that the limit presented in \citet{holoien16b} is an order of magnitude higher than actually implied by the non-detection of this source. From our re-examination of the data, we derive an upper limit of $5.8\times10^{-2}$ counts/sec in the 0.3--10.0 keV energy band. Assuming a power law component with $\Gamma=1.75$, similar to that of known AGN \citep[e.g.,][]{tozzi06,marchesi16, liu17, ricci17}, we derive an unabsorbed flux limit of $1.9\times10^{-12}$ ergs cm$^{-2}$ s$^{-1}$, which corresponds to $L_{X}\sim10^{43}$ erg s$^{-1}$ at the distance of the TDE. This is comparable to the observed X-ray luminosities of AGN found in deep extragalactic surveys, so we cannot rule out the presence of an underlying AGN. However, in \citet{holoien16b} we highlighted that the host galaxy has a mid-IR color from WISE of $W_{1}-W_{2}=0.06\pm0.06$, which implies that any AGN activity in the host galaxy is not strong \citep{assef13}. Further long-term observations are needed to confirm whether the hard component of the X-ray emission is associated with the TDE itself and is possible evidence of a jet, or whether it persists well after the TDE fades, which would imply the presence of an underlying, weak AGN.


\begin{figure}
\centering
\subfloat{{\includegraphics[width=0.45\textwidth]{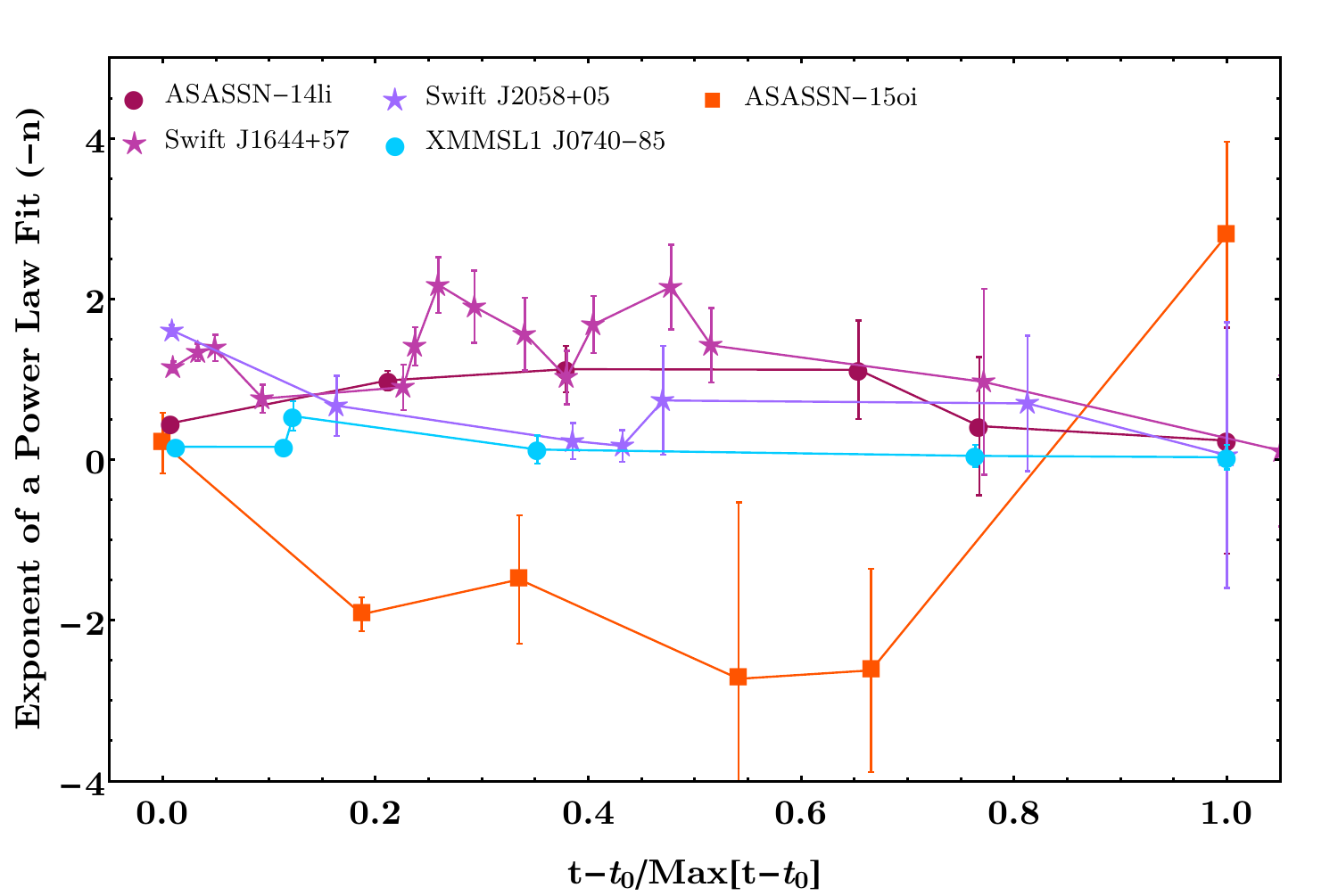}}}
\caption{Best-fit exponent of a power law fit (-n) and its uncertainty as the time of discovery/peak ($t_{0}$) goes to infinity, as a function of $t-t_{0}$ for ASASSN-15oi (orange squares). Overlaid are the evolution of the jetted events X-ray TDEs Swift J1644+57 and Swift J2058+05 (pink and purple stars, respectively), and the non-jetted X-ray TDEs ASASSN-14li and XMMSL1 J0740-85 (burgundy and cyan circles, respectively) as presented in \citet{auchettl18}. The power law index of \oi\ exhibits a similar magnitude of variation to those seen in the other X-ray TDEs, rather than the large-scale variations seen in AGN, though it is of the opposite sign.}
\label{fig:powerlaw}
\end{figure}

In order to investigate the origin of the X-ray emission, we compare properties of the X-ray emission of \oi\ to those of other X-ray emitting TDEs using the metric presented by \citet[][]{auchettl18} to distinguish TDEs from AGN. \citet{auchettl18} fit a temporal power law, $f\propto(t-t_0)^{-n}$, to the X-ray emission of several TDEs and AGN and found that TDEs show a relatively coherent decay with $n\sim0-2$, while AGN varied dramatically on short timescales with a much broader range of indices ($-10<n<15$). Figure~\ref{fig:powerlaw} shows $-n$ as a function of time for \oi\ and several X-ray TDEs from the analysis of \citet{auchettl18}. While ASASSN-15oi shows a constant power-law exponent, unlike AGN which have n values that vary dramatically over short time-scales, it is unique in that it has a negative value of n, due to its remarkable power-law brightening in the X-rays with time. As such, it is very likely that this thermal component arises from the TDE itself.


\begin{figure}
\centering
\includegraphics[width=0.45\textwidth]{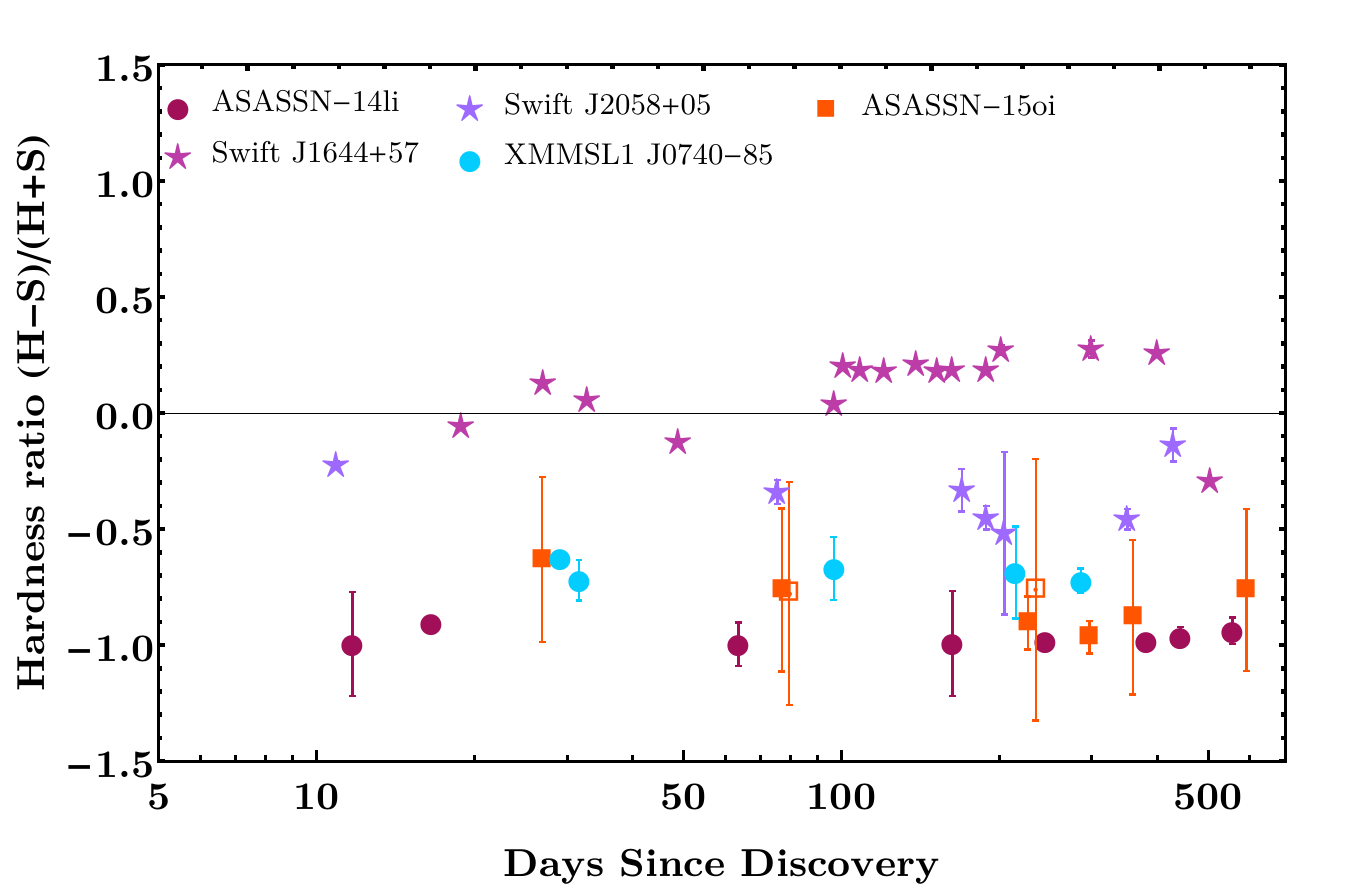}
\caption{X-ray hardness ratio as a function of time for ASASSN-15oi. Here the hardness ratio derived from the merged Swift observations are plotted as filled orange squares, while the ratio derived from the two deep XMM-Newton observations are shown as unfilled orange squares. Overlaid are the hardness ratios for the non-jetted thermal X-ray TDEs (ASASSN-14li and XMMSL1 J0740-85) and the jetted X-ray TDEs (Swift J1644+57, Swift J2058+05), taken from \citet{auchettl17}.}
\label{fig:hardness_ratio}
\end{figure} 

In Figure \ref{fig:hardness_ratio} we show the evolution of the hardness ratio (HR) as compared to the ``X-ray TDEs'' from \citet{auchettl18}. Similar to what was seen in XMMSL1 J0740-85, the X-ray emission of ASASSN-15oi shows little color evolution with time and is relatively soft. Here we have assumed that the hard X-ray component is not associated with the TDE.

However, if one assumes that the hard X-ray component is associated with the TDE itself, rather than an underlying AGN, ASASSN-15oi would exhibit strong variation in its HR, much larger than that seen from the non-jetted thermal and jetted X-ray TDEs plotted in Figure \ref{fig:hardness_ratio}. Under this assumption, we find that the HR derived from \textit{XMM-Newton} (unfilled squares) would be closer to HR$\sim$0, rather than HR$\sim$-0.7 presented in the plot. If the hard X-ray component is associated with the TDE itself, in the form of a possible jet, then this variation may be indicative of variations in the accretion rate of this source, assuming that the jet luminosity is related to the accretion rate. However, this variation is much larger than that seen from other X-ray TDEs that exhibit a jet. In addition, this change in HR would be in conflict with the HR derived from the \textit{Swift} observations (filled squares) that were taken nearly simultaneously, which would imply very large variability on a very short timescale. While similarly strong variability on short timescales has been seen in AGN \citep[e.g.,][]{auchettl18}, this would be highly unique for a TDE. Irrespective of the origin of the underlying hard X-ray component, this highlights the importance of performing deep observations using high sensitivity X-ray instruments such as \textit{XMM-Newton} as a complement to those observations taken using \textit{Swift}.


\begin{figure*}
\begin{minipage}{\textwidth}
\centering
\subfloat{{\includegraphics[width=0.95\textwidth]{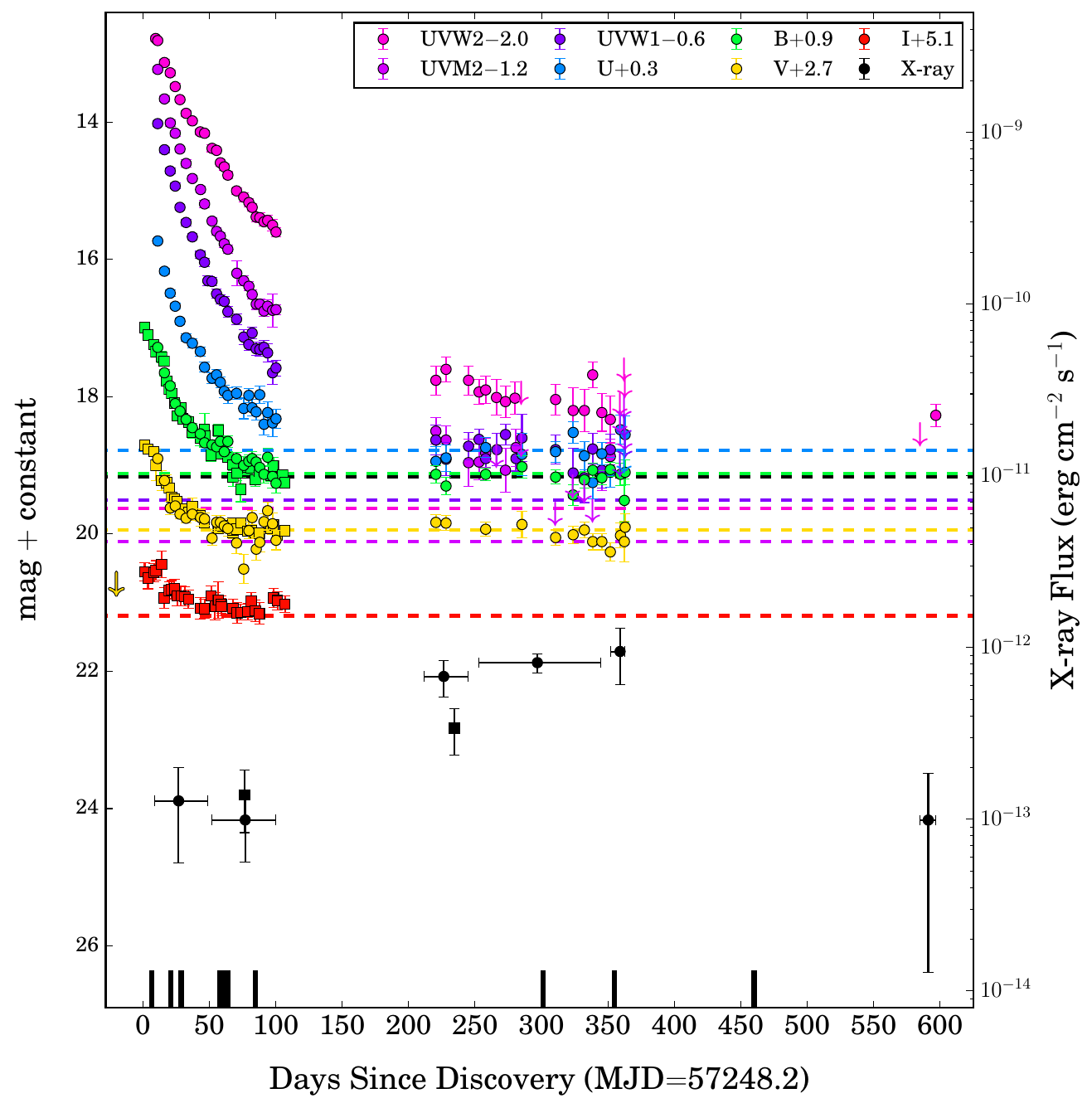}}}
\caption{Evolution of {\oi} in the X-ray, UV, and optical bands from discovery to $\sim600$ days after discovery. Follow-up data obtained from {\swift} (X-ray, UV, and optical) are shown as circles and follow-up data obtained by the Las Cumbres Observatory 1-m telescopes (optical) and \textit{XMM-Newton} (X-ray) are shown as squares. Upper limits are indicated with downward arrows, and the $V$-band upper limit at $-20$ days is a 3$\sigma$ upper limit from ASAS-SN. All optical and UV magnitudes are on the Vega system (left scale), and unabsorbed X-ray fluxes in the $0.3-10.0$ keV energy range are shown in ergs~cm$^{-2}$~s$^{-1}$ (right scale), and the two scales have the same dynamic range. UV/optical data have not been corrected for Galactic extinction. The XRT data have been binned to increase signal-to-noise of the detections, and horizontal error bars indicate the date range for each bin. Horizontal dashed lines show the synthesized UVOT and $BVI$ host magnitudes and the archival X-ray upper limit from ROSAT, and vertical marks along the time-axis show the dates of our spectroscopic observations.}
\label{fig:lightcurve}
\end{minipage}
\end{figure*}


\subsection{Photometric Evolution}
\label{sec:phot_anal}

Figure~\ref{fig:lightcurve} shows the photometric evolution of \oi\ in the X-ray, UV, and optical, extending the results from \citet{holoien16b} to $\sim600$ days after discovery. As in \citet{holoien16b}, the \swift\ $B$- and $V$-band data were converted to the Johnson-Cousins system to be consistent with data obtained from the Las Cumbres Observatory 1-m telescopes. The colored horizontal dashed lines represent the synthesized host magnitudes in the UV and optical bands, and the horizontal black line shows the X-ray upper limit from ROSAT. 

Two main features of the photometric evolution stand out. First, the X-ray emission shows significant evolution at later times, as discussed in \S\ref{sec:xray_anal}. Second, while the spectral features faded quickly, the UV emission from the TDE leveled off at around 250 days after discovery, and continues to be brighter than the host 600 days after discovery. This is similar to ASASSN-14li, which also had a similar leveling off in its late-time UV emission \citep{brown17a}. This may indicate a transition in the physical mechanism responsible for the emission.

To characterize this excess UV emission, we corrected the UV and optical magnitudes for Galactic extinction assuming $R_V=3.1$ and $A_V=0.19$ \citep{schlafly11}, deriving the extinction in other filters using a \citet{cardelli89} extinction law. We then subtracted the host flux in each band and modeled the UV and optical SED of \oi\ as a blackbody using Markov Chain Monte Carlo methods, as was done for the previous ASAS-SN TDEs \citep{holoien14b,holoien16a,brown16a,brown17a}. When performing the fit, we excluded any data where $f_\lambda/f_{\lambda, host}<0.5$, as we did in \citet{holoien16b}. Many of the later epochs of \swift\ observations resulted in non-detections or marginal detections, and in order to increase the significance of our flux measurements we binned the data taken 200 days or more after discovery in 20-day bins. These binned fluxes were used to fit the blackbody SED at late-times, rather than the individual observations. We find that in all epochs the UV and optical emission is well described by a blackbody, although in some cases data was only available in one or two filters.


\begin{figure}
\centering
\subfloat{{\includegraphics[width=0.45\textwidth]{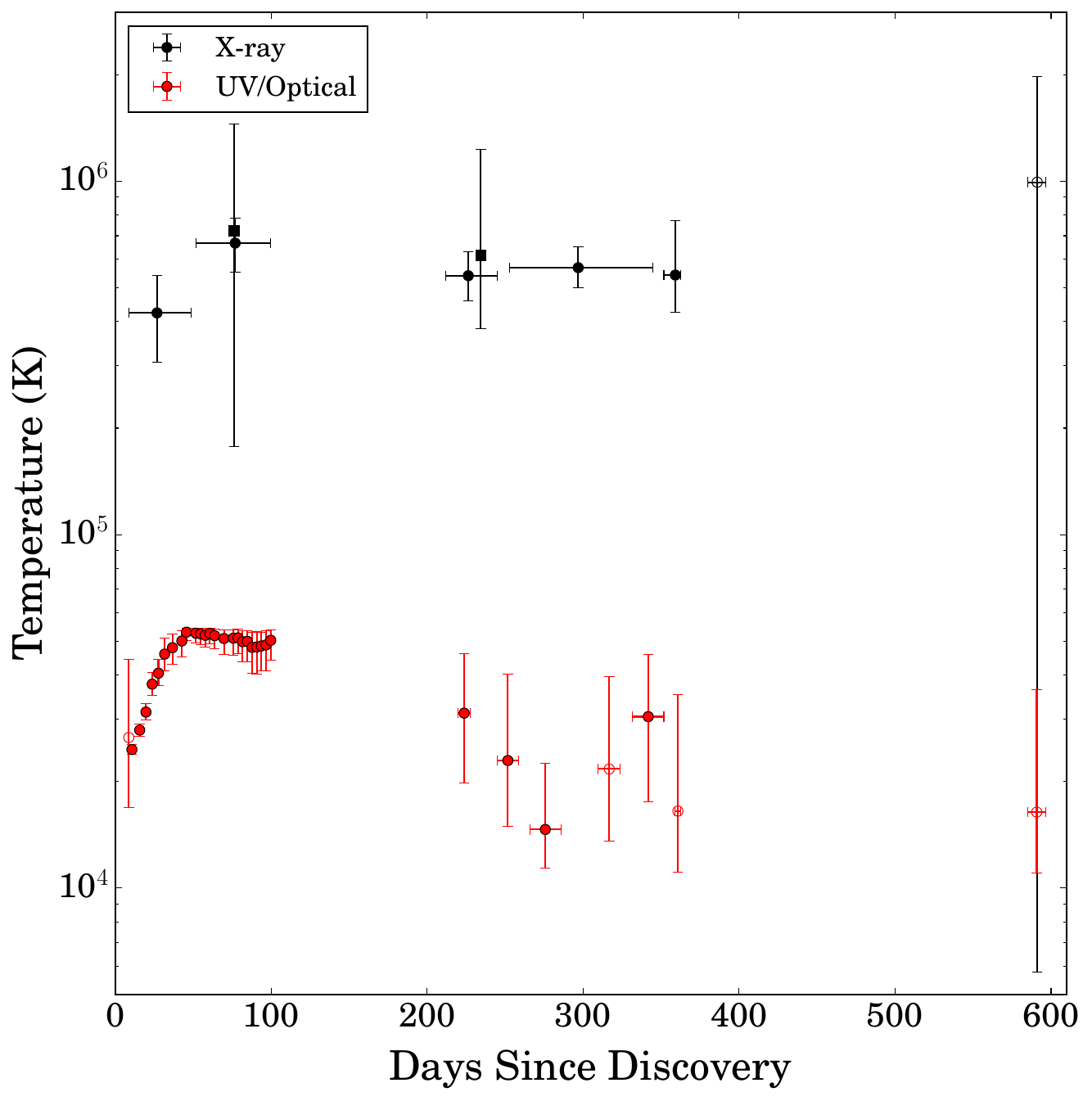}}}
\caption{Temperature evolution of \oi\ from blackbody fits to the UV/optical SED (red points) and of the thermal component of the X-ray emission from \swift\ XRT (black circles) and \textit{XMM-Newton} (black squares). Data obtained more than 200 days after discovery were binned, with the horizontal bars indicating the date range of the observations. Filled circles indicate bins with 3 or more epochs of observation, while open circles have 1 or 2 epochs of observation.}
\label{fig:temp_evol}
\end{figure}

In \citet{holoien16b} we found that when we fit the SED, the fits preferred a rising temperature in early epochs. Because of this, we fit the SED using a fairly restrictive temperature prior: for epochs within 10 days of discovery, the temperature was fit with a prior of $T=2\times10^4$~K, and for later epochs, we fit the data with a prior of $T=(2+(\Delta t-10)/2)\times10^4$~K, where $\Delta t=MJD-57248.2$ and the temperature is capped at $T=40000$~K, with a $\log$ uncertainty of $\pm0.05$ dex in all epochs. In the later epochs, the source is becoming redder, indicating a decreasing temperature, so for this manuscript we revised the blackbody fits using a much less restrictive prior of $10000$~K~$\leq T \leq55000$~K in all epochs. Figure~\ref{fig:temp_evol} shows the blackbody temperature evolution of \oi\ using this new prior. In the earliest epochs, the temperature is not well-constrained because our \swift\ data do not span the peak of the SED, and we find that the fits prefer to fit values near the higher end of our prior range. However, in the late-time data, the data prefer cooler temperatures of $20000$~K$-30000$~K, indicating that the flare cooled before leveling off.

Figure~\ref{fig:temp_evol} also shows the temperature evolution of the thermal component of the X-ray data. The X-ray emission from \oi\ can be well described using a simple blackbody, assuming that the hard X-ray tail is not associated with the TDE. As originally highlighted in \citet{holoien16b}, the initial temperature evolution of ASASSN-15oi is unique compared to other well-characterized TDEs such as ASASSN-14li \citep{brown17a}. There is some hint that the X-ray temperature follows a similar behavior to that seen in the UV/optical, with the X-ray temperature initially increasing and then flattening. However, the large uncertainty on the blackbody temperature derived from the first \textit{XMM-Newton} observation makes it difficult to make strong statements about the initial evolution. In all epochs, the X-ray temperature is an order of magnitude or more hotter than the UV/optical component, suggesting the emission arises from a different region around the SMBH.

\begin{figure}
\centering
\subfloat{{\includegraphics[width=0.45\textwidth]{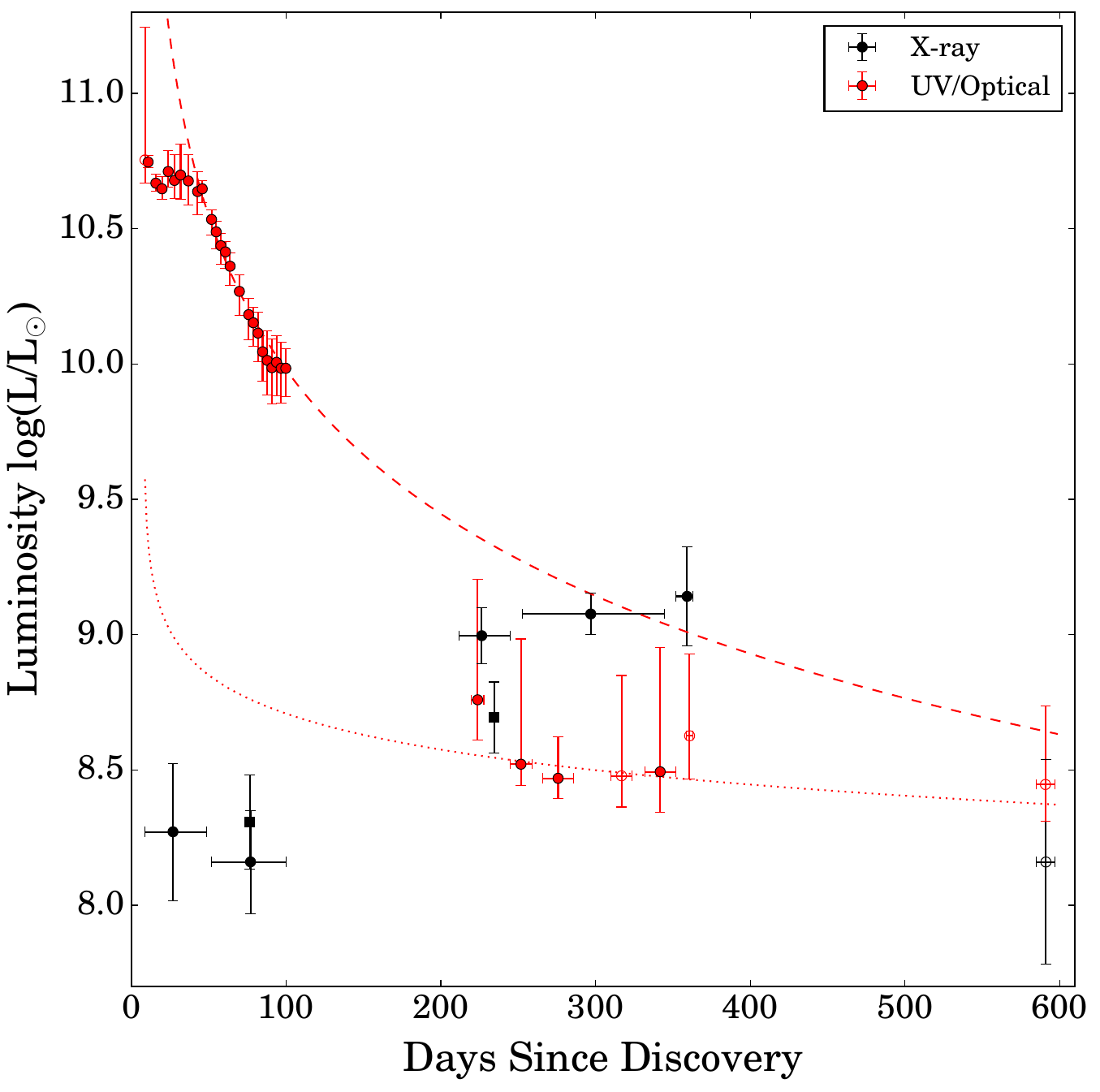}}}
\caption{UV/optical luminosity evolution of \oi\ from the blackbody SED fits (red circles) and X-ray luminosity evolution from \swift\ (black circles) and \textit{XMM-Newton} (black squares). The red dashed line shows a $t^{-5/3}$ power-law fit to the early UV/optical data and the red dotted line shows a $t^{-5/12}$ power-law fit to the late-time UV/optical data. When fitting the late-time data, we fixed the time of disruption $t_0$ to be the same as that fit from the early data.}
\label{fig:lum_evol}
\end{figure}

Figure~\ref{fig:lum_evol} shows the UV/optical and X-ray luminosity evolution of \oi. The early observations show a plateau in the first couple of weeks followed by a steady decline in luminosity, possibly indicating the transient was discovered near its peak luminosity. Conversely, the late-time data indicate that the UV/optical luminosity was approximately constant for the last $\sim400$ days of observation. In the Figure we also show $t^{-5/3}$ power-law fits to the early data (dashed lines) and $t^{-5/12}$ power-law fits to the late-time data (dotted lines). We find that the early data is reasonably well-fit by a $t^{-5/3}$ power-law profile, as we found in \citet{holoien16b}, but that this profile cannot fit the late-time data. The late-time data are reasonably well-fit by a $t^{-5/12}$ power-law profile, though the best-fit power law index is poorly constrained and highly sensitive to the choice of $t_0$. For the late-time fit shown in the Figure, we fix $t_0$ to the value fit from the early-time data. When the bolometric luminosity is fallback-dominated we expect a $t^{-5/3}$ decline, and when it is dominated by emission from the accretion disk we expect a $t^{-5/12}$ decline \citep{lodato11}, though the decline rates in individual UV/optical filters can differ from the bolometric decline rate due to band effects. While the best-fit power laws to the early and late-time data are not exactly $t^{-5/3}$ and $t^{-5/12}$, the differing power law fits indicate that we may be observing a transition from a fallback-dominated regime to a disk-dominated regime, or some other transition in the emission mechanism like that seen in other TDEs \citep[see e.g.,][]{auchettl17}.

The combined blackbody-plus-power-law X-ray luminosity is also shown in Figure~\ref{fig:lum_evol}. At early times, the X-ray emission is 2 orders of magnitude or more weaker than the UV/optical emission, but during the period of increased X-ray emission between $\sim200$ days and $\sim400$ days, the X-rays become the dominant source of emission. The X-ray emission fades again to early-time levels in the last epoch of observation at $\sim600$ days, but the UV/optical emission has also faded significantly by that time, and the X-rays and UV/optical make similar contributions to the overall luminosity. Thus, while the X-rays are relatively insignificant at early times, they represent a significant or dominant fraction of the energy emitted at later times.


Integrating the combined optical/UV and X-ray luminosity curves gives a total radiated energy of $E=(1.32\pm0.06)\times10^{51}$~ergs for the $\sim600$~days of follow-up observations. This required accretion of only $\Delta M\sim7.4\times10^{-3}\eta_{0.1}^{-1}$~{\msun} of mass, where $\eta_{0.1}=0.1\eta$ is the radiative efficiency, to power the event, assuming the flare is powered by accretion. This is similar to the amount of mass needed to power ASASSN-14li, $\Delta M\sim4\times10^{-3}\eta_{0.1}^{-1}$~{\msun}, as derived by \citet{brown17a}.

\begin{figure}
\centering
\subfloat{{\includegraphics[width=0.45\textwidth]{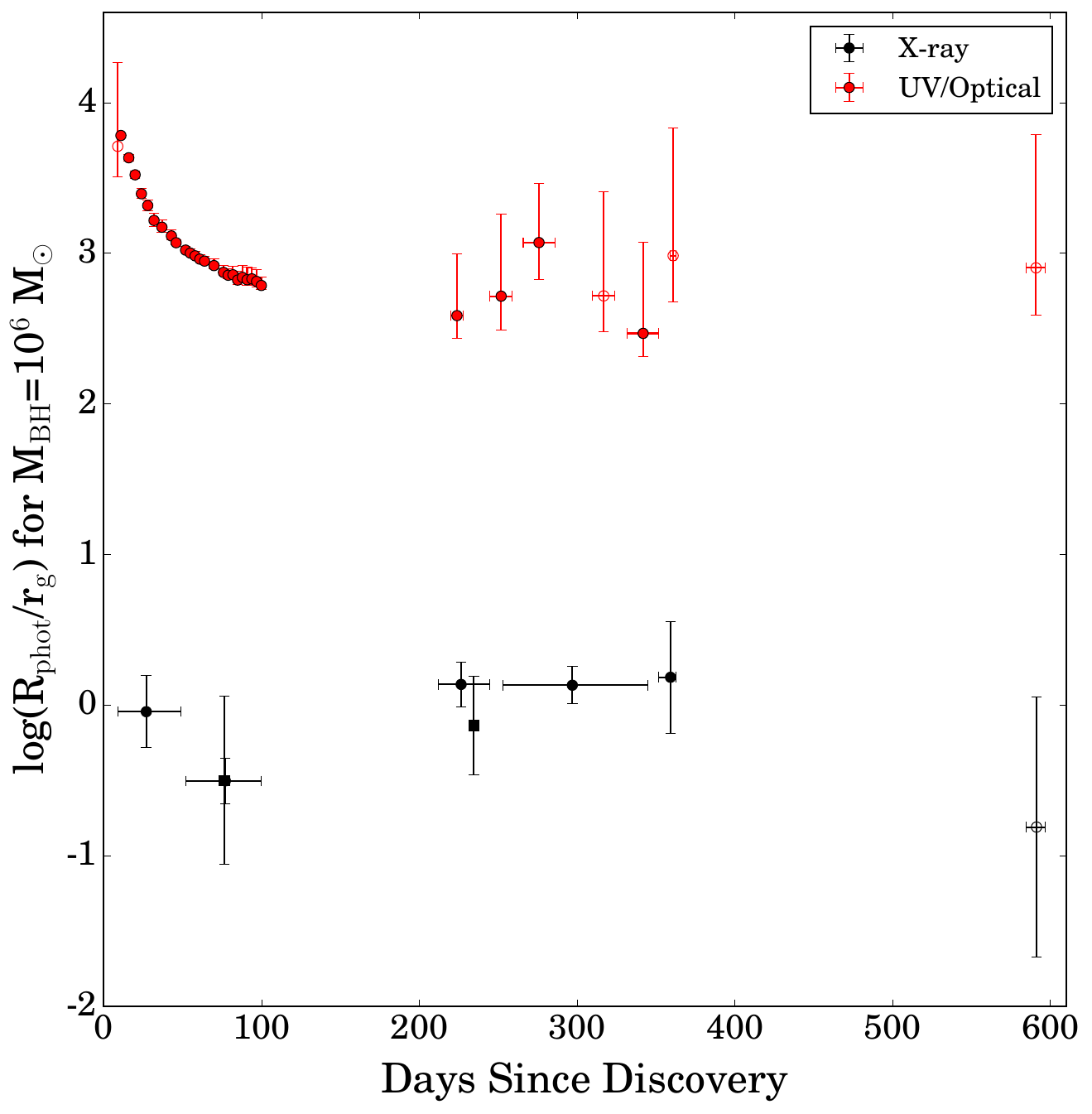}}}
\caption{Evolution of the photospheric radius derived from the blackbody SED fits (red circles) and X-ray emission from \swift\ (black circles) and \textit{XMM-Newton} (black squares), shown in units of the gravitational radius of a $10^6$~\msun\ black hole. Similar to what was seen in ASASSN-14li, the size of the UV/optical emitting region shrinks and then levels off at later times, though \oi\ exhibits a smaller UV/optical photosphere than ASASSN-14li. Conversely, the size of the X-ray emitting region remains relatively constant throughout the event.}
\label{fig:rad_evol}
\end{figure}

The evolution of the effective radius for the UV/optical and X-ray emission are shown in Figure~\ref{fig:rad_evol}. We originally estimated the mass of the SMBH to be $10^{7.1}$~\msun\ in \citet{holoien16b} using the host galaxy bulge mass and the bulge-black hole mass relation from \citet{mcconnell13}. However, more recently \citet{gezari17} and \citet{mockler18} found that the mass of the SMBH in \host\ is more likely on the order of $M_h\sim10^{6.4}$~{\msun}, using a total stellar mass-black hole mass relation \citep{reines15} and the Modular Open Source Fitter for Transients fitting code \citep{guillochon17}, respectively. Because of these new, lower mass estimates, we scale the photospheric radii shown in Figure~\ref{fig:rad_evol} to the gravitational radius of a $10^6$~\msun\ mass black hole.


As was seen with ASASSN-14li \citep{brown17a}, the radius implied by the UV/optical emission initially shrinks and then levels off at later times. The blackbody radius of \oi\ is larger than that of ASASSN-14li early in the flare ($R\sim10^4$ gravitational radii, compared to $R\sim10^3$ gravitational radii for ASASSN-14li), but both TDEs level off at a similar size of roughly $10^3$ gravitational radii.

We also show the characteristic radius of the X-ray emission in Figure~\ref{fig:rad_evol}, again assuming that a spherical blackbody well describes the soft X-ray emission. The X-ray emitting surface is approximately 1 gravitational radius in size, again assuming a $10^{6}$~{\msun} black hole. This is significantly smaller than the characteristic radius of the UV/optical emission from ASASSN-15oi. Unlike the  UV/optical emission, we do not see a decrease in the size of the photosphere with time. The size of the X-ray emitting region seems to remain relatively constant, although the uncertainties are fairly large. We also observed little evolution in ASASSN-14li \citep{brown17a}, and similar to the UV/optical emitting surface, the emitting X-ray surface of \oi\ is again comparable to that of ASASSN-14li. There is some evidence that the size of the X-ray emitting region increases between 200--400 days, and it is possible the increase in the observed X-ray flux of the blackbody component may be a natural result of this.

\section{Discussion}
\label{sec:disc}

After discovering the nearby TDE \oi\ and publishing our initial observations in \citet{holoien16b}, we continued to monitor this object in the optical, UV, and X-rays to roughly 600 days after discovery. Our late-time observations show that while \oi\ shows many similarities with other TDEs, it is also unique in a number of ways.

Our spectroscopic monitoring program indicates that the spectroscopic features, including the strong blue continuum and emission lines, faded rapidly, and were essentially gone roughly 2 months post-discovery. Such rapid evolution of the spectral features was only seen in one other optically discovered TDE, iPTF16fnl, which was significantly less luminous than \oi\ \citep{blagorodnova17,brown17b}. The host-subtracted spectra indicate that the continuum slope is reasonably approximated by the Raleigh-Jeans tail of a blackbody, and that the emission lines narrow as they fade, as has been seen in other optically discovered TDEs \citep[e.g.,][]{holoien14b,holoien16a,brown16a,brown17a}. The host-subtracted spectra also make it apparent that there is likely a weak H$\alpha$ emission feature, though the helium emission lines are significantly stronger, and thus \oi\ still falls on the He-rich end of the continuum of spectroscopic features described in \citet{arcavi14}.

As first reported by \citet{gezari17} the evolution of the X-ray emission seen in \oi\ is unique. In particular, the X-ray emission brightens by roughly an order of magnitude by $\sim250$~days before fading back to early levels in the latest epochs of observation. Our analyzes indicate that the bulk of this evolution comes from the thermal component of the X-rays, and that the hard component, when detected, is weaker and relatively constant throughout the flare. While all the detected X-ray emission is below an archival limit from ROSAT, the X-ray emission evolves without significant short-timescale variations, is relatively soft in nature and shows little variation in hardness ratio with time, consistent with that of other X-ray TDEs \citep{auchettl18}. We conclude that the thermal component of the X-ray emission is almost certainly due to the TDE, while the origin of the hard component is still unclear. Though the lack of evolution in the hard component indicates it is likely related to the host, rather than the TDE, further X-ray observations showing that the hard component remains constant are needed to definitively show that this is the case.

The TDE also continues to emit in the UV at 600 days, remaining well above the estimated host flux despite leveling off and showing very little evolution in the late-time observations. This behavior is very similar to that of ASASSN-14li \citep{brown17a}, but ASASSN-14li also showed long-lasting optical spectroscopic features. \oi\ shows significant temperature evolution, growing much hotter in early epochs before cooling and remaining constant during the late-time observations. The early luminosity evolution is relatively well-fit by a $t^{-5/3}$ power law decline while the late-time data can be fit by a $t^{-5/12}$ decline, indicating we may be seeing a transition from a fallback-dominated emission regime to a disk-dominated regime. Finally, the blackbody radius associated with the UV and optical emission is orders of magnitude larger than that of the thermal X-rays, indicating that the two types of radiation are likely coming from different regions around the SMBH.

ASASSN-15oi provides another piece of evidence that detailed late-time observations of TDEs are necessary to fully characterize the emission of these rare transients. Very little evolution was seen in the X-ray emission at early times, but the late-time observations showed that \oi\ exhibited unique evolution that may help shed light on the physical processes responsible for this emission. Further, while the spectral features faded rapidly in \oi, it continued to exhibit UV emission for hundreds of days after discovery, indicating again that transient TDE emission can contaminate host galaxies for years following a tidal disruption. The late-time UV observations are once again able to rule out a purely $t^{-5/3}$ decline in the luminosity, indicating that the evolution of the luminosity associated with TDEs is more complex than simple theoretical models. Continued study of \oi\ will help shed light on the nature of the non-thermal X-ray emission, and future TDE discoveries should be observed for similarly long periods of time in order to fully characterize the emission from these events and allow us to more accurately model the physical processes responsible.

\section*{Acknowledgments}

The authors thank K. Stanek, E. Ramirez-Ruiz, and B. Mockler for their suggestions and comments for improving the manuscript. We thank N. Morrell for obtaining follow-up observations. We thank Las Cumbres Observatory and its staff for their continued support of ASAS-SN.

ASAS-SN is supported by NSF grant AST-1515927. Development of ASAS-SN has been supported by NSF grant AST-0908816, the Center for Cosmology and AstroParticle Physics at the Ohio State University, the Mt. Cuba Astronomical Foundation, and by George Skestos.

CSK is supported by NSF grant AST-1515876. Support for JLP is in part provided by the Ministry of Economy, Development, and Tourism's Millennium Science Initiative through grant IC120009, awarded to The Millennium Institute of Astrophysics, MAS.

This research has made use of the XRT Data Analysis Software (XRTDAS) developed under the responsibility of the ASI Science Data Center (ASDC), Italy. At Penn State the NASA {\swift} program is support through contract NAS5-00136.

This research was made possible through the use of the AAVSO Photometric All-Sky Survey (APASS), funded by the Robert Martin Ayers Sciences Fund.

This research has made use of data provided by Astrometry.net \citep{barron08}.

This paper uses data products produced by the OIR Telescope Data Center, supported by the Smithsonian Astrophysical Observatory.

This publication makes use of data products from the Two Micron All Sky Survey, which is a joint project of the University of Massachusetts and the Infrared Processing and Analysis Center/California Institute of Technology, funded by the National Aeronautics and Space Administration and the National Science Foundation.

This publication makes use of data products from the Wide-field Infrared Survey Explorer, which is a joint project of the University of California, Los Angeles, and the Jet Propulsion Laboratory/California Institute of Technology, funded by the National Aeronautics and Space Administration.

This research has made use of the NASA/IPAC Extragalactic Database (NED), which is operated by the Jet Propulsion Laboratory, California Institute of Technology, under contract with the National Aeronautics and Space Administration.

This work is based in part on observations collected at the European Organisation for Astronomical Research in the Southern Hemisphere, Chile as part of PESSTO, (the Public ESO Spectroscopic Survey for Transient Objects Survey) ESO program 188.D-3003, 191.D-0935.

This publication makes use of data obtained from the Weizmann interactive supernova data repository \citep[WISEREP;][]{yaron12}.

\bibliography{bibliography.bib}
\bsp	


\appendix
\section{Follow-up Photometry}
All UV and optical follow-up photometry are presented in Table~\ref{tab:phot} in the Vega system. 


\begin{table*}
\begin{minipage}{\textwidth}
\centering
\caption{Photometric data of \oi.\hfill}
\renewcommand{\arraystretch}{1.2}
\begin{tabular}{cccc|cccc}
\hline
MJD & Magnitude &  Filter & Telescope & MJD & Magnitude &  Filter & Telescope\\
\hline
57249.532 & 15.45 0.129 & $I$ & LCO & 57321.794 & 17.14 0.077 & $V$ & LCO\\ 
57252.002 & 15.54 0.158 & $I$ & LCO & 57327.129 & 17.26 0.045 & $V$ & LCO\\ 
57256.428 & 15.46 0.133 & $I$ & LCO & 57329.844 & 17.25 0.055 & $V$ & LCO\\ 
57258.008 & 15.43 0.143 & $I$ & LCO & 57332.791 & 17.28 0.051 & $V$ & LCO\\ 
57262.251 & 15.34 0.193 & $I$ & LCO & 57335.816 & 17.29 0.065 & $V$ & LCO\\ 
57264.083 & 15.83 0.150 & $I$ & LCO & 57344.415 & 17.22 0.060 & $V$ & LCO\\ 
57268.052 & 15.72 0.129 & $I$ & LCO & 57346.420 & 17.17 0.066 & $V$ & LCO\\ 
57269.908 & 15.71 0.133 & $I$ & LCO & 57349.423 & 17.32 0.056 & $V$ & LCO\\ 
57271.882 & 15.70 0.143 & $I$ & LCO & 57355.035 & 17.25 0.052 & $V$ & LCO\\ 
57273.924 & 15.80 0.148 & $I$ & LCO & 57264.392 & 16.52 0.061 & $V$ & {\swift}\\ 
57276.874 & 15.80 0.142 & $I$ & LCO & 57268.788 & 16.92 0.100 & $V$ & {\swift}\\ 
57279.895 & 15.81 0.140 & $I$ & LCO & 57272.514 & 16.89 0.071 & $V$ & {\swift}\\ 
57282.476 & 15.85 0.133 & $I$ & LCO & 57276.096 & 17.01 0.071 & $V$ & {\swift}\\ 
57291.383 & 15.98 0.154 & $I$ & LCO & 57280.688 & 17.07 0.071 & $V$ & {\swift}\\ 
57294.522 & 15.99 0.141 & $I$ & LCO & 57285.415 & 17.01 0.091 & $V$ & {\swift}\\ 
57299.495 & 15.80 0.139 & $I$ & LCO & 57291.590 & 17.06 0.130 & $V$ & {\swift}\\ 
57302.481 & 15.95 0.200 & $I$ & LCO & 57294.588 & 17.08 0.091 & $V$ & {\swift}\\ 
57304.921 & 15.86 0.262 & $I$ & LCO & 57300.236 & 17.36 0.100 & $V$ & {\swift}\\ 
57307.106 & 15.93 0.126 & $I$ & LCO & 57303.628 & 17.13 0.100 & $V$ & {\swift}\\ 
57307.140 & 15.96 0.125 & $I$ & LCO & 57306.494 & 17.13 0.100 & $V$ & {\swift}\\ 
57316.018 & 15.98 0.134 & $I$ & LCO & 57309.413 & 17.18 0.100 & $V$ & {\swift}\\ 
57319.059 & 16.05 0.152 & $I$ & LCO & 57312.137 & 17.22 0.110 & $V$ & {\swift}\\ 
57321.796 & 16.04 0.116 & $I$ & LCO & 57318.671 & 17.43 0.150 & $V$ & {\swift}\\ 
57327.133 & 16.03 0.126 & $I$ & LCO & 57324.051 & 17.81 0.210 & $V$ & {\swift}\\ 
57329.849 & 15.88 0.128 & $I$ & LCO & 57327.961 & 17.25 0.120 & $V$ & {\swift}\\ 
57332.796 & 16.02 0.159 & $I$ & LCO & 57330.420 & 17.06 0.100 & $V$ & {\swift}\\ 
57335.821 & 16.06 0.157 & $I$ & LCO & 57333.480 & 17.52 0.160 & $V$ & {\swift}\\ 
57346.424 & 15.83 0.140 & $I$ & LCO & 57336.141 & 17.42 0.140 & $V$ & {\swift}\\ 
57349.419 & 15.87 0.133 & $I$ & LCO & 57339.334 & 17.12 0.120 & $V$ & {\swift}\\ 
57355.039 & 15.92 0.121 & $I$ & LCO & 57342.069 & 16.96 0.130 & $V$ & {\swift}\\ 
57249.533 & 16.01 0.053 & $V$ & LCO & 57345.793 & 17.15 0.160 & $V$ & {\swift}\\ 
57252.004 & 16.06 0.055 & $V$ & LCO & 57348.384 & 17.39 0.140 & $V$ & {\swift}\\ 
57256.429 & 16.10 0.053 & $V$ & LCO & 57468.704 & 17.13 0.110 & $V$ & {\swift}\\ 
57258.009 & 16.30 0.225 & $V$ & LCO & 57476.354 & 17.14 0.110 & $V$ & {\swift}\\ 
57259.334 & 16.20 0.051 & $V$ & {\swift} & 57506.200 & 17.23 0.100 & $V$ & {\swift}\\ 
57262.252 & 16.52 0.144 & $V$ & LCO & 57533.509 & 17.16 0.190 & $V$ & {\swift}\\ 
57266.065 & 16.55 0.112 & $V$ & LCO & 57558.821 & 17.35 0.110 & $V$ & {\swift}\\ 
57268.053 & 16.63 0.047 & $V$ & LCO & 57572.057 & 17.31 0.130 & $V$ & {\swift}\\ 
57269.909 & 16.76 0.050 & $V$ & LCO & 57580.496 & 17.24 0.120 & $V$ & {\swift}\\ 
57271.883 & 16.79 0.062 & $V$ & LCO & 57586.741 & 17.41 0.120 & $V$ & {\swift}\\ 
57273.926 & 16.83 0.058 & $V$ & LCO & 57593.652 & 17.41 0.120 & $V$ & {\swift}\\ 
57276.875 & 16.93 0.057 & $V$ & LCO & 57600.036 & 17.56 0.130 & $V$ & {\swift}\\ 
57278.172 & 16.94 0.062 & $V$ & LCO & 57607.674 & 17.32 0.180 & $V$ & {\swift}\\ 
57282.475 & 16.98 0.050 & $V$ & LCO & 57610.592 & 17.41 0.300 & $V$ & {\swift}\\ 
57285.424 & 16.90 0.130 & $V$ & LCO & 57611.184 & 17.20 0.200 & $V$ & {\swift}\\ 
57291.382 & 17.03 0.081 & $V$ & LCO & 57249.531 & 16.09 0.055 & $B$ & LCO\\ 
57294.521 & 17.13 0.074 & $V$ & LCO & 57252.001 & 16.20 0.075 & $B$ & LCO\\ 
57304.919 & 17.22 0.074 & $V$ & LCO & 57256.426 & 16.34 0.042 & $B$ & LCO\\ 
57307.104 & 17.18 0.047 & $V$ & LCO & 57258.007 & 16.44 0.065 & $B$ & LCO\\ 
57307.138 & 17.18 0.056 & $V$ & LCO & 57262.250 & 16.52 0.082 & $B$ & LCO\\ 
57312.170 & 17.20 0.068 & $V$ & LCO & 57264.081 & 16.58 0.062 & $B$ & LCO\\ 
57315.877 & 17.28 0.082 & $V$ & LCO & 57266.062 & 16.87 0.063 & $B$ & LCO\\ 
57315.894 & 17.14 0.073 & $V$ & LCO & 57268.051 & 16.99 0.068 & $B$ & LCO\\ 
57316.016 & 17.24 0.075 & $V$ & LCO & 57269.906 & 17.05 0.060 & $B$ & LCO\\ 
57319.057 & 17.23 0.059 & $V$ & LCO & 57271.880 & 17.19 0.074 & $B$ & LCO\\ 
\hline
\end{tabular}
\label{tab:phot}
\end{minipage}
\end{table*}

\begin{table*}
\begin{minipage}{\textwidth}
\centering
\renewcommand{\arraystretch}{1.2}
\begin{tabular}{cccc|cccc}
\hline
MJD & Magnitude &  Filter & Telescope & MJD & Magnitude &  Filter & Telescope\\
\hline
57273.923 & 17.38 0.067 & $B$ & LCO & 57580.493 & 18.31 0.122 & $B$ & {\swift}\\ 
57276.872 & 17.26 0.069 & $B$ & LCO & 57586.737 & 18.17 0.102 & $B$ & {\swift}\\ 
57278.169 & 17.42 0.076 & $B$ & LCO & 57593.648 & 18.28 0.102 & $B$ & {\swift}\\ 
57282.474 & 17.47 0.065 & $B$ & LCO & 57600.033 & 18.16 0.102 & $B$ & {\swift}\\ 
57285.423 & 17.62 0.083 & $B$ & LCO & 57607.671 & 18.23 0.161 & $B$ & {\swift}\\ 
57291.381 & 17.70 0.115 & $B$ & LCO & 57610.590 & 18.61 0.312 & $B$ & {\swift}\\ 
57294.520 & 17.58 0.242 & $B$ & LCO & 57611.181 & 18.20 0.181 & $B$ & {\swift}\\ 
57295.130 & 17.67 0.085 & $B$ & LCO & 57259.328 & 15.43 0.036 & $U$ & {\swift}\\ 
57299.493 & 17.95 0.098 & $B$ & LCO & 57264.383 & 15.87 0.036 & $U$ & {\swift}\\ 
57304.918 & 17.59 0.086 & $B$ & LCO & 57268.722 & 16.19 0.046 & $U$ & {\swift}\\ 
57307.137 & 17.93 0.067 & $B$ & LCO & 57272.506 & 16.38 0.045 & $U$ & {\swift}\\ 
57312.168 & 17.98 0.077 & $B$ & LCO & 57276.088 & 16.60 0.045 & $U$ & {\swift}\\ 
57315.873 & 18.27 0.130 & $B$ & LCO & 57280.678 & 16.84 0.054 & $U$ & {\swift}\\ 
57315.891 & 18.09 0.122 & $B$ & LCO & 57285.412 & 16.92 0.063 & $U$ & {\swift}\\ 
57316.015 & 18.09 0.086 & $B$ & LCO & 57291.387 & 17.04 0.063 & $U$ & {\swift}\\ 
57319.055 & 18.21 0.108 & $B$ & LCO & 57294.581 & 17.27 0.073 & $U$ & {\swift}\\ 
57321.792 & 18.45 0.179 & $B$ & LCO & 57300.229 & 17.43 0.082 & $U$ & {\swift}\\ 
57327.124 & 18.12 0.058 & $B$ & LCO & 57303.621 & 17.38 0.082 & $U$ & {\swift}\\ 
57329.839 & 18.14 0.049 & $B$ & LCO & 57306.486 & 17.49 0.082 & $U$ & {\swift}\\ 
57332.786 & 18.31 0.099 & $B$ & LCO & 57309.408 & 17.62 0.092 & $U$ & {\swift}\\ 
57335.812 & 18.18 0.102 & $B$ & LCO & 57312.131 & 17.68 0.112 & $U$ & {\swift}\\ 
57344.411 & 18.25 0.077 & $B$ & LCO & 57318.667 & 17.65 0.092 & $U$ & {\swift}\\ 
57346.415 & 18.11 0.131 & $B$ & LCO & 57324.048 & 17.87 0.151 & $U$ & {\swift}\\ 
57349.414 & 18.27 0.076 & $B$ & LCO & 57327.910 & 17.68 0.102 & $U$ & {\swift}\\ 
57353.420 & 18.24 0.064 & $B$ & LCO & 57330.418 & 17.86 0.122 & $U$ & {\swift}\\ 
57355.030 & 18.35 0.071 & $B$ & LCO & 57333.413 & 17.92 0.122 & $U$ & {\swift}\\ 
57259.329 & 16.38 0.036 & $B$ & {\swift} & 57336.134 & 17.67 0.122 & $U$ & {\swift}\\ 
57264.384 & 16.75 0.045 & $B$ & {\swift} & 57339.331 & 18.10 0.152 & $U$ & {\swift}\\ 
57268.724 & 16.94 0.045 & $B$ & {\swift} & 57342.066 & 17.93 0.162 & $U$ & {\swift}\\ 
57272.507 & 17.20 0.045 & $B$ & {\swift} & 57345.787 & 18.08 0.202 & $U$ & {\swift}\\ 
57276.089 & 17.31 0.054 & $B$ & {\swift} & 57348.377 & 18.02 0.141 & $U$ & {\swift}\\ 
57280.679 & 17.43 0.054 & $B$ & {\swift} & 57468.701 & 18.64 0.215 & $U$ & {\swift}\\ 
57285.413 & 17.55 0.063 & $B$ & {\swift} & 57476.350 & 18.59 0.194 & $U$ & {\swift}\\ 
57291.388 & 17.64 0.063 & $B$ & {\swift} & 57506.198 & 18.44 0.142 & $U$ & {\swift}\\ 
57294.582 & 17.77 0.073 & $B$ & {\swift} & 57533.506 & 18.54 0.305 & $U$ & {\swift}\\ 
57300.230 & 17.80 0.073 & $B$ & {\swift} & 57558.815 & 18.50 0.152 & $U$ & {\swift}\\ 
57303.622 & 17.83 0.082 & $B$ & {\swift} & 57572.053 & 18.22 0.162 & $U$ & {\swift}\\ 
57306.487 & 17.75 0.073 & $B$ & {\swift} & 57580.493 & 18.56 0.203 & $U$ & {\swift}\\ 
57309.409 & 17.90 0.082 & $B$ & {\swift} & 57586.736 & 18.95 0.235 & $U$ & {\swift}\\ 
57312.132 & 17.75 0.082 & $B$ & {\swift} & 57593.648 & 18.53 0.162 & $U$ & {\swift}\\ 
57318.668 & 18.00 0.092 & $B$ & {\swift} & 57600.033 & 18.8 0.213 & $U$ & {\swift}\\ 
57324.049 & 18.10 0.132 & $B$ & {\swift} & 57607.670 & 18.18 0.212 & $U$ & {\swift}\\ 
57327.911 & 18.04 0.092 & $B$ & {\swift} & 57610.590 & $>$18.45 & $U$ & {\swift}\\ 
57330.418 & 18.00 0.102 & $B$ & {\swift} & 57611.181 & 18.25 0.273 & $U$ & {\swift}\\ 
57333.414 & 18.05 0.102 & $B$ & {\swift} & 57259.325 & 14.62 0.042 & $W1$ & {\swift}\\ 
57336.135 & 18.13 0.132 & $B$ & {\swift} & 57264.379 & 15.00 0.042 & $W1$ & {\swift}\\ 
57339.332 & 18.23 0.132 & $B$ & {\swift} & 57268.719 & 15.31 0.042 & $W1$ & {\swift}\\ 
57342.066 & 17.98 0.132 & $B$ & {\swift} & 57272.503 & 15.53 0.042 & $W1$ & {\swift}\\ 
57345.788 & 18.26 0.171 & $B$ & {\swift} & 57276.085 & 15.84 0.042 & $W1$ & {\swift}\\ 
57348.378 & 18.36 0.141 & $B$ & {\swift} & 57280.675 & 16.06 0.05 & $W1$ & {\swift}\\ 
57468.701 & 18.23 0.122 & $B$ & {\swift} & 57285.411 & 16.27 0.05 & $W1$ & {\swift}\\ 
57476.350 & 18.40 0.122 & $B$ & {\swift} & 57291.384 & 16.53 0.05 & $W1$ & {\swift}\\ 
57506.198 & 18.23 0.092 & $B$ & {\swift} & 57294.579 & 16.64 0.058 & $W1$ & {\swift}\\ 
57533.506 & 18.12 0.161 & $B$ & {\swift} & 57297.048 & 16.91 0.076 & $W1$ & {\swift}\\ 
57558.816 & 18.27 0.102 & $B$ & {\swift} & 57300.226 & 16.92 0.058 & $W1$ & {\swift}\\ 
57572.054 & 18.53 0.151 & $B$ & {\swift} & 57303.618 & 17.10 0.067 & $W1$ & {\swift}\\  
\hline
\end{tabular}
\end{minipage}
\end{table*}

\begin{table*}
\begin{minipage}{\textwidth}
\centering
\renewcommand{\arraystretch}{1.2}
\begin{tabular}{cccc|cccc}
\hline
MJD & Magnitude &  Filter & Telescope & MJD & Magnitude &  Filter & Telescope\\
\hline
57306.483 & 17.18 0.067 & $W1$ & {\swift} & 57476.354 & 19.83 0.202 & $M2$ & {\swift}\\ 
57309.407 & 17.21 0.067 & $W1$ & {\swift} & 57493.299 & 20.16 0.341 & $M2$ & {\swift}\\ 
57312.128 & 17.36 0.076 & $W1$ & {\swift} & 57501.204 & 20.15 0.262 & $M2$ & {\swift}\\ 
57318.666 & 17.47 0.076 & $W1$ & {\swift} & 57506.200 & 20.02 0.212 & $M2$ & {\swift}\\ 
57324.047 & 17.73 0.104 & $W1$ & {\swift} & 57514.309 & $>$19.87  & $M2$ & {\swift}\\ 
57327.909 & 17.84 0.085 & $W1$ & {\swift} & 57521.082 & 20.27 0.321 & $M2$ & {\swift}\\ 
57330.416 & 17.67 0.085 & $W1$ & {\swift} & 57528.985 & 19.94 0.272 & $M2$ & {\swift}\\ 
57333.412 & 17.90 0.095 & $W1$ & {\swift} & 57533.510 & $>$19.72  & $M2$ & {\swift}\\ 
57336.132 & 17.91 0.104 & $W1$ & {\swift} & 57558.822 & $>$20.72  & $M2$ & {\swift}\\ 
57339.330 & 17.88 0.104 & $W1$ & {\swift} & 57572.057 & $>$20.32  & $M2$ & {\swift}\\ 
57342.064 & 17.96 0.133 & $W1$ & {\swift} & 57580.497 & $>$20.39  & $M2$ & {\swift}\\ 
57345.784 & 18.25 0.173 & $W1$ & {\swift} & 57586.741 & $>$20.67  & $M2$ & {\swift}\\ 
57348.374 & 18.18 0.114 & $W1$ & {\swift} & 57593.652 & 20.28 0.282 & $M2$ & {\swift}\\ 
57468.700 & 19.23 0.222 & $W1$ & {\swift} & 57600.036 & 20.07 0.252 & $M2$ & {\swift}\\ 
57476.348 & 19.51 0.242 & $W1$ & {\swift} & 57607.675 & $>$19.37  & $M2$ & {\swift}\\ 
57493.302 & 19.32 0.202 & $W1$ & {\swift} & 57610.592 & $>$19.16  & $M2$ & {\swift}\\ 
57501.209 & 19.22 0.143 & $W1$ & {\swift} & 57611.184 & $>$19.63  & $M2$ & {\swift}\\ 
57506.197 & 19.50 0.212 & $W1$ & {\swift} & 57257.794 & 14.78 0.036 & $W2$ & {\swift}\\ 
57514.700 & 19.37 0.232 & $W1$ & {\swift} & 57259.330 & 14.81 0.042 & $W2$ & {\swift}\\ 
57521.086 & 19.15 0.153 & $W1$ & {\swift} & 57264.386 & 15.13 0.042 & $W2$ & {\swift}\\ 
57528.989 & 19.50 0.182 & $W1$ & {\swift} & 57268.725 & 15.28 0.042 & $W2$ & {\swift}\\ 
57533.504 & 19.20 0.341 & $W1$ & {\swift} & 57272.508 & 15.48 0.042 & $W2$ & {\swift}\\ 
57558.812 & 19.37 0.212 & $W1$ & {\swift} & 57276.091 & 15.67 0.042 & $W2$ & {\swift}\\ 
57572.051 & 19.71 0.361 & $W1$ & {\swift} & 57280.681 & 15.87 0.042 & $W2$ & {\swift}\\ 
57580.491 & 19.78 0.361 & $W1$ & {\swift} & 57285.413 & 15.98 0.042 & $W2$ & {\swift}\\ 
57586.735 & 19.36 0.222 & $W1$ & {\swift} & 57291.389 & 16.14 0.042 & $W2$ & {\swift}\\ 
57593.646 & 19.67 0.262 & $W1$ & {\swift} & 57294.583 & 16.16 0.042 & $W2$ & {\swift}\\ 
57600.031 & 19.37 0.222 & $W1$ & {\swift} & 57300.231 & 16.38 0.050 & $W2$ & {\swift}\\ 
57607.668 & $>$19.41  & $W1$ & {\swift} & 57303.623 & 16.41 0.050 & $W2$ & {\swift}\\ 
57610.589 & $>$18.84  & $W1$ & {\swift} & 57306.488 & 16.59 0.050 & $W2$ & {\swift}\\ 
57611.179 & $>$19.14  & $W1$ & {\swift} & 57309.410 & 16.65 0.050 & $W2$ & {\swift}\\ 
57259.335 & 14.43 0.042 & $M2$ & {\swift} & 57312.133 & 16.77 0.050 & $W2$ & {\swift}\\ 
57264.394 & 14.86 0.042 & $M2$ & {\swift} & 57318.668 & 17.00 0.050 & $W2$ & {\swift}\\ 
57268.789 & 15.21 0.050 & $M2$ & {\swift} & 57324.049 & 17.09 0.067 & $W2$ & {\swift}\\ 
57272.515 & 15.36 0.042 & $M2$ & {\swift} & 57327.911 & 17.17 0.058 & $W2$ & {\swift}\\ 
57276.098 & 15.59 0.042 & $M2$ & {\swift} & 57330.418 & 17.24 0.058 & $W2$ & {\swift}\\ 
57280.689 & 15.80 0.050 & $M2$ & {\swift} & 57333.414 & 17.38 0.067 & $W2$ & {\swift}\\ 
57285.416 & 16.02 0.042 & $M2$ & {\swift} & 57336.136 & 17.39 0.067 & $W2$ & {\swift}\\ 
57291.591 & 16.18 0.050 & $M2$ & {\swift} & 57339.332 & 17.45 0.067 & $W2$ & {\swift}\\ 
57294.589 & 16.39 0.050 & $M2$ & {\swift} & 57342.067 & 17.43 0.076 & $W2$ & {\swift}\\ 
57300.237 & 16.64 0.050 & $M2$ & {\swift} & 57345.789 & 17.50 0.085 & $W2$ & {\swift}\\ 
57303.629 & 16.79 0.050 & $M2$ & {\swift} & 57348.379 & 17.60 0.067 & $W2$ & {\swift}\\ 
57306.495 & 16.86 0.050 & $M2$ & {\swift} & 57468.702 & 19.76 0.212 & $W2$ & {\swift}\\ 
57309.413 & 16.97 0.050 & $M2$ & {\swift} & 57476.351 & 19.60 0.182 & $W2$ & {\swift}\\ 
57312.138 & 17.05 0.058 & $M2$ & {\swift} & 57493.296 & 19.76 0.212 & $W2$ & {\swift}\\ 
57318.869 & 17.40 0.182 & $M2$ & {\swift} & 57501.199 & 19.93 0.182 & $W2$ & {\swift}\\ 
57324.052 & 17.51 0.076 & $M2$ & {\swift} & 57506.198 & 19.90 0.202 & $W2$ & {\swift}\\ 
57327.962 & 17.59 0.067 & $M2$ & {\swift} & 57514.307 & 20.01 0.262 & $W2$ & {\swift}\\ 
57330.421 & 17.71 0.067 & $M2$ & {\swift} & 57521.079 & 20.07 0.232 & $W2$ & {\swift}\\ 
57333.481 & 17.85 0.076 & $M2$ & {\swift} & 57528.257 & 20.02 0.232 & $W2$ & {\swift}\\ 
57336.142 & 17.85 0.076 & $M2$ & {\swift} & 57533.507 & $>$19.75  & $W2$ & {\swift}\\ 
57339.334 & 17.95 0.076 & $M2$ & {\swift} & 57558.817 & 20.04 0.222 & $W2$ & {\swift}\\ 
57342.069 & 17.88 0.095 & $M2$ & {\swift} & 57572.054 & 20.20 0.331 & $W2$ & {\swift}\\ 
57345.794 & 17.94 0.242 & $M2$ & {\swift} & 57580.494 & 20.20 0.311 & $W2$ & {\swift}\\ 
57348.385 & 17.93 0.076 & $M2$ & {\swift} & 57586.738 & 19.68 0.192 & $W2$ & {\swift}\\ 
57468.704 & 19.70 0.192 & $M2$ & {\swift} & 57593.649 & 20.23 0.272 & $W2$ & {\swift}\\ 
\hline
\end{tabular}
\end{minipage}
\end{table*}

\begin{table*}
\begin{minipage}{\textwidth}
\centering
\renewcommand{\arraystretch}{1.2}
\begin{tabular}{cccc|cccc}
\hline
MJD & Magnitude &  Filter & Telescope & MJD & Magnitude &  Filter & Telescope\\
\hline
57600.034 & 20.33 0.341 & $W2$ & {\swift} & 57611.182 & $>$19.66  & $W2$ & {\swift}\\ 
57607.671 & $>$19.92  & $W2$ & {\swift} & 57833.642 & $>$20.36  & $W2$ & {\swift}\\ 
57610.590 & $>$19.41  & $W2$ & {\swift} & 57845.002 & 20.27 0.163 & $W2$ & {\swift}\\ 
\hline
\end{tabular}
\\
\medskip
\raggedright
\noindent All magnitudes and uncertainties are presented in the Vega system. Uncertainties are given next to the magnitude measurements, except for upper limits. Data are not corrected for Galactic extinction. The abbreviation ``LCO'' refers to the Las Cumbres Observatory telescopes.
\end{minipage}
\end{table*}

\end{document}